\pdfoutput=1 
\documentclass{egpubl}

\JournalPaper         
 \electronicVersion 

\PrintedOrElectronic

\usepackage{t1enc,dfadobe}
\usepackage{egweblnk}
\usepackage{cite}

\usepackage{empheq}
\usepackage{graphicx}
\usepackage{amsmath}
\usepackage[lined, ruled]{algorithm2e}
\usepackage{color}
\usepackage{listings}
\usepackage{soul,xcolor} 

\usepackage{epstopdf} 

\usepackage{mathtools}
\newcommand{\defeq}{\vcentcolon=}

\newcommand*\euler{\mathrm{e}}
\newcommand{\albedo}{\alpha}
\newcommand{\vect}[1]{\mathbf{#1}}

\newcommand{\ud}{\,\mathrm{d}} 

\usepackage{datetime}

\usepackage{SIunits}

\usepackage{verbatim}

\usepackage[normalem]{ulem}

\usepackage{subfigure}
\title{Flux-Limited Diffusion for Multiple Scattering in Participating Media \vspace{-0.85cm}}

\author[D. Koerner, J. Portsmouth,  F. Sadlo, T. Ertl, \& B. Eberhardt]{D. Koerner$^1$, J. Portsmouth\thanks{Joint first author}$^2$,  F. Sadlo$^1$, T. Ertl$^1$, and B. Eberhardt$^3$  \vspace{-0.15cm} \\ 
         $^1$University of Stuttgart, Stuttgart, Germany \quad $^2$The Moving Picture Company \quad $^3$Stuttgart Media University
}

\volume{33}   
\issue{2}     

\begin{document}

\maketitle

\begin{abstract}
For the rendering of multiple scattering effects in participating media, methods based on the diffusion approximation are an extremely efficient alternative to Monte Carlo path tracing. However, in sufficiently transparent regions, classical diffusion approximation suffers from non-physical radiative fluxes which leads to a poor match to correct light transport. In particular, this prevents the application of classical diffusion approximation to heterogeneous media, where opaque material is embedded within transparent regions. To address this limitation, we introduce flux-limited diffusion, a technique from the astrophysics domain. This method provides a better approximation to light transport than classical diffusion approximation, particularly when applied to heterogeneous media, and hence broadens the applicability of diffusion-based techniques. We provide an algorithm for flux-limited diffusion, which is validated using the transport theory for a point light source in an infinite homogeneous medium. We further demonstrate that our implementation of flux-limited diffusion produces more accurate renderings of multiple scattering in various heterogeneous datasets than classical diffusion approximation, by comparing both methods to ground truth renderings obtained via volumetric path tracing.


\end{abstract}

\section{Introduction}

The rendering of multiply-scattered light is highly computationally expensive in the presence of dense participating media due to the proliferation of scattering paths.
A brute force solution exists in the form of volumetric path tracing, in which the random scattering of photons is explicitly simulated via a Monte Carlo method such as introduced by Lafortune et al.~\cite{lafortune1996rendering}. In practice, however, these methods are computationally too demanding for many applications, since the calculation becomes increasingly expensive as the medium becomes more opaque and the mean free path (the local average distance travelled by a photon) between scattering events decreases. Eventually path tracing becomes intractable, and we enter a diffusive regime where photons undergo long random walks in the medium. In this regime, the classical diffusion equation provides a very good approximation to multiply-scattered light, which is vastly more efficient than path tracing. It is a simple partial differential equation which can be solved efficiently via a variety of methods, both analytical and numerical.

Diffusion methods are thus clearly of great interest as a means of accelerating the
simulation of
multiple scattering in high opacity media. 
However, application of these methods to more general volumetric media, such as
clouds in which opaque regions are embedded in relatively transparent regions, has been problematic for the basic reason that the classical diffusion approximation (CDA) becomes increasingly inaccurate as the medium becomes more transparent. In fact, the classical diffusion equation for the light field becomes singular in the limit of a zero extinction ``vacuum'' region, as elaborated on in Section~\ref{sec:fld}. In the context of subsurface scattering~\cite{jensen2001practical}, this is not an issue because the vacuum outside the medium is dealt with analytically by imposing boundary conditions at the surface. However, in a general heterogeneous volume this is not possible as there may not exist a well-defined boundary. So it has not been clear thus far how to extend the diffusion approach to these more general situations.

\begin{figure*}[!t]
\hspace*{\fill}%
  \subfigure[Path tracing (\unit{23}{\minute})\label{subfig:clouds-mc}]{\includegraphics[width=0.3\linewidth]{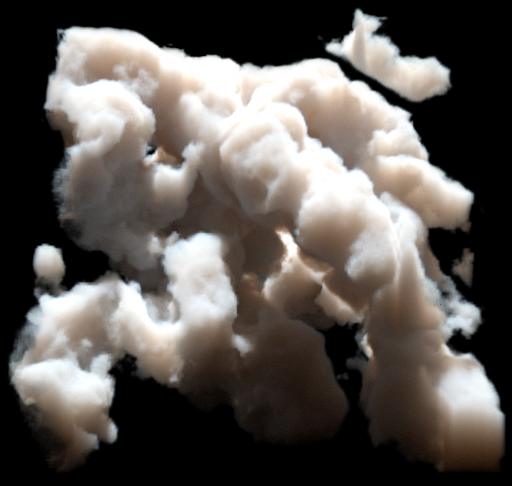}}
\hspace{1.5pt}
  \subfigure[Flux-limited diffusion (\unit{0.6}{\second})\label{subfig:clouds-fld}]{\includegraphics[width=0.3\linewidth]{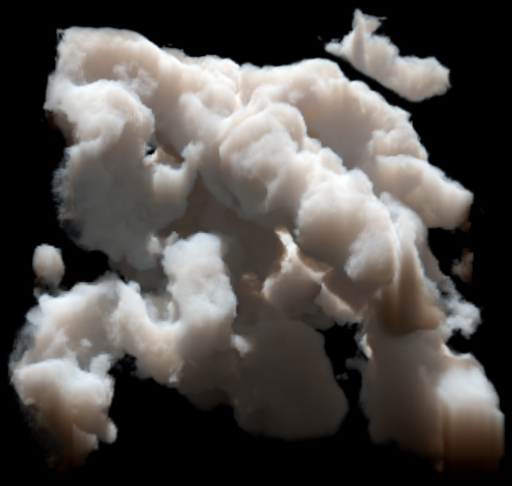}}
\hspace{1.5pt}
  \subfigure[Diffusion approximation (\unit{0.4}{\second})\label{subfig:clouds-da}]{\includegraphics[width=0.3\linewidth]{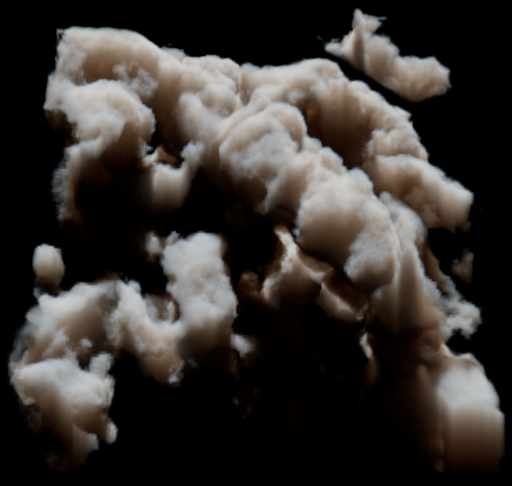}}
 \hspace*{\fill}%
  \caption{
Renderings of a heterogeneous volume which contains regions of high and low opacity, embedded in vacuum. 
This is problematic for the classical diffusion approximation~\subref{subfig:clouds-da}, 
which becomes invalid near and within regions of low opacity.
Our approach employing flux-limited diffusion~\subref{subfig:clouds-fld} addresses this deficiency and as a result provides a better approximation to volumetric path tracing~\subref{subfig:clouds-mc} at a fraction of the cost.
}
  \label{fig:nebulae}
\end{figure*}

Our contribution is the introduction of the flux-limited diffusion (FLD) method to the graphics community and its application to the problem of rendering participating media with multiply-scattered light. This method is a standard radiative transfer technique in other fields (in particular, astrophysics \cite{Levermore_Pomraning_1981, 2001ApJS..135...95T} and nuclear physics \cite{1974JMP....15...75L}) and its underlying theory was first fully explored by Levermore and Pomraning\cite{Levermore_Pomraning_1981}. FLD was developed specifically to produce a better approximation to multiple scattering in the low optical depth regime than CDA. We will explain how a grid-based diffusion solver can be easily extended to FLD by modifying the diffusion coefficient, and demonstrate that FLD in the mentioned cases performs significantly better than CDA, in the sense that it results in images closer to ground truth at affordable additional computational cost (see Figure~\ref{fig:nebulae}). Beyond that, we believe that flux-limited diffusion has the potential to improve other  techniques for multiple scattering in participating media which are based on diffusion approximation, such as \cite{Arbree:2011:HSS:1990770.1990989, Zhang:2013:FGI:2448196.2448205}.

In the next two sections, we discuss related work followed by a brief review of CDA theory. In Section~\ref{sec:fld}, we introduce the theory of FLD and give some intuition for its physical basis. We show how to extend an existing CDA solver to FLD in Section~\ref{sec:numerical_method}. In Section~\ref{sec:results} we evaluate our method using the analytical solution for a point source as a testbed, then render various datasets using FLD and compare to ground truth images obtained via volumetric path tracing. 

\section{Related Work}

Multiple scattering in participating media has been extensively studied and we restrict the discussion to the work which is most related to ours. For a broader overview, we refer to 
Cerezo et al.~\cite{CEREZO:2005:INRIA-00510151:1}.

The theory of multiple scattering as a diffusion process has a long history in the radiative transfer literature \cite{ishimaru1978wave, Lightman1986}. In computer graphics, the theory of light diffusion has been most profitably applied to the rendering of subsurface scattering (SSS). This is the searchlight problem of finding the exitant radiance under incident illumination at the boundary of a homogeneous scattering medium. Jensen et al. \cite{jensen2001practical} pioneered the application of the radiative diffusion approximation to this problem, solving the diffusion equation with appropriate boundary conditions imposed at the surface. Further work has refined this technique \cite{d2011quantized, yan2012accurate, habel13pbd}. This line of research cannot be applied to our problem since SSS deals with media with well-defined surfaces, while we are interested in finding the radiance \emph{within} heterogeneous volumes which contain regions of low opacity or vacuum.

Wang et al.~\cite{wang2008modeling} extended SSS to a heterogeneous translucent volumetric medium modeled as an irregular finite element grid
on which CDA was solved. Arbree et al.~\cite{Arbree:2011:HSS:1990770.1990989} later in similar approaches addressed various problems in Wang et al.'s work and simplified discretization by using a tetrahedral mesh basis. 
Li et al.~\cite{Li2013} used the same tetrahedral representation but a more efficient method for constructing the coefficient matrix, which allowed realtime application under changing mesh topology.    
In contrast to our approach, all this work still focuses on rendering highly opaque media with well defined surfaces.

CDA was first applied to the multiple scattering of light in heterogeneous participating media defined on a grid by Stam~\cite{stam1995multiple}, refining earlier work by~\cite{Kajiya:1984:RTV:800031.808594}. Here, however, the breakdown of CDA within more transparent regions was not accommodated for.
In applying CDA to the rendering of clouds, Max et al.~\cite{DBLP:conf/wscg/MaxSMIN04} recognized the problem, and to avoid it, introduced a hybrid scheme in which diffusion is applied to discrete homogenous cloud cores with high optical depth and well-defined boundaries on a grid. Ray tracing is used to propagate light across the empty space between grids.

Other related grid-based approaches to scattering in inhomogeneous participating media include Light Propagation Volumes \cite{Kaplanyan:2010:CLP:1730804.1730821, Borlum:2011:SSL:2018323.2018325}, and Lattice Boltzmann methods \cite{Geist:2004:LL:2383533.2383582, Qiu:2007:LVG:1313046.1313126}. These are similar to the diffusion methods in the sense that light is propagated between cells on a grid.
Bouthors et al. \cite{Bouthors:2008:IMA:1342250.1342277} described a semi-analytical approach to multiple scattering in media bounded by geometry in which a pre-computed path tracing solution in slab geometry is used.
All these methods are less physically-based than ours, using phenomenological models rather than direct approximation of the radiative transfer equation.  
Another general approach based directly on the radiative transfer equation (RTE)~\cite{chandrasekhar1960radiative} is Light Propagation Maps \cite{Fattal:2009:PMI:1477926.1477933}. This is essentially a formulation of the classic Discrete Ordinate Method~\cite{chandrasekhar1960radiative} in which the light field in each grid cell is discretized into ray direction bins. This is a potentially more accurate method than the diffusion approximation, but also much more computationally expensive. 

\section{Classical Diffusion Approximation}
\label{sec:cda}

In this section we review CDA as a preliminary to our introduction of FLD in Section~\ref{sec:fld}. For more details, we refer the reader to~\cite{ishimaru1978wave} and~\cite{stam1995multiple}.
In Table~\ref{Nomenclature} we give a summary of the symbols used. 

We assume that the properties of the medium are specified, in the form of an absorption coefficient field $\sigma_a(\mathbf{x})$ and scattering coefficient field $\sigma_s(\mathbf{x})$, both with units of inverse length. We can also express these in terms of the total extinction field $\sigma_t(\mathbf{x}) = \sigma_a(\mathbf{x}) +\sigma_s(\mathbf{x})$ and the dimensionless scattering albedo $\alpha(\mathbf{x}) = \sigma_s(\mathbf{x})/\sigma_t(\mathbf{x})$. The mean free path is given by $\sigma_t(\mathbf{x})^{-1}$. We also allow for the possibility that the medium generates an isotropic radiance field $j(\mathbf{x})$ (with units of radiated power per unit volume) due to self-emission. We specialize in our whole treatment in this paper to
the case of isotropic scattering and discuss anisotropy in Section~\ref{sec:conclusion}.

Given these medium properties, the radiance field $L(\mathbf{x}, \boldsymbol{\omega})$, for all points $\mathbf{x}$ in the volume and all ray directions $\boldsymbol{\omega}$, is in principle determined by solving the RTE, an integral equation which the radiance field must satisfy
\begin{equation} \label{rte}
(\mathbf{\omega}\cdot\nabla) L(\mathbf{x}, \mathbf{\omega}) = -\sigma_t(\mathbf{x}) L(\mathbf{x}, \mathbf{\omega}) + Q(\mathbf{x}) \ ,
\end{equation}
with 
\begin{equation} \label{rte2}
Q(\mathbf{x}) =  \frac{j(\mathbf{x})}{4\pi} + \sigma_s\int_{\Omega} \frac{1}{4\pi}L(\mathbf{x}, \mathbf{\omega}')\ud\omega' \ .
\end{equation}
The source term $Q(\mathbf{x})$ represents self-emission and in-scattering (the factor $4\pi$ in the first term is the conversion factor from power density to power density per unit steradian, see \cite{Lightman1986}). 

Our particular task in volume rendering is to determine the radiance $L(\mathbf{x}_e, \boldsymbol{\omega})$ at the endpoint $\mathbf{x}_e$ of each primary ray of light propagating in direction $\boldsymbol{\omega}$ towards the camera through the volume. 
The total radiance at the endpoint $\mathbf{x}_e$ is the sum of the attenuated ``back-light'', and a term due to the light added by the medium at each sample along the ray (attenuated by the transmittance between the sample point $\mathbf{x}_n$ and $\mathbf{x}_e$):
\begin{equation} \label{primary_raymarch}
L(\mathbf{x}_e, \boldsymbol{\omega}) = L(\mathbf{x}_s, \boldsymbol{\omega})  T(\mathbf{x}_s, \mathbf{x}_e) + \Delta x \,\sum_n  T(\mathbf{x}_n, \mathbf{x}_e) Q(\mathbf{x}_n) \ .
\end{equation}
The transmittance function $T(\mathbf{x}_a, \mathbf{x}_b) = \exp{(-\tau(\mathbf{x}_a, \mathbf{x}_b))}$ gives the attenuation of radiance due to absorption and out-scattering along a given ray segment with endpoints $\mathbf{x}_a$ and $\mathbf{x}_b$ and optical depth (number of mean-free-paths) $\tau=\int_{\mathbf{x}_a}^{\mathbf{x}_b}\sigma_t(\mathbf{x}) \ud\mathbf{x}$. This is computed numerically by breaking the ray into segments. With segment endpoints $\mathbf{x}_n$, and segment length $\Delta x$, we have in discretized form $T(\mathbf{x}_s, \mathbf{x}_e) \approx \exp\left( - \sum_n \sigma_t(\mathbf{x}_n) \Delta x\right)$.

\begin{table}
\small
    \caption{Nomenclature and units\label{Nomenclature}}
    \centering
\vspace{-0.2cm}
    \begin{tabular}{cl} 
    \\ \hline\hline
    Symbol                                                    & Meaning 						\\ \hline
    ~ $\sigma_s$ & scattering coefficient $[ \mathrm{m}^{-1}]$ \\
    ~ $\sigma_a$ & absorption coefficient $[\mathrm{m}^{-1}]$ \\
    ~ $\sigma_t=\sigma_a+\sigma_s$ & extinction coefficient $[\mathrm{m}^{-1}]$\\
    $\alpha=\sigma_s/\sigma_t$ & albedo \\
    $L(\mathbf{x}, \boldsymbol{\omega})$ & radiance field $[\mathrm{Wm}^{-2}\mathrm{sr}^{-1}]$\\
    $j(\vect{x})$ & self-emission field $[\mathrm{Wm}^{-3}]$\\
    $Q(\vect{x})$ & source term $[\mathrm{Wm}^{-3}\mathrm{sr}^{-1}]$\\
    $L_{ri}(\vect{x})$ & reduced incident radiance $[\mathrm{Wm}^{-2}]$\\
    $L_m(\vect{x})$ & medium radiance $[\mathrm{Wm}^{-2}]$\\
    $\phi(\vect{x})$ & diffuse fluence $[\mathrm{Wm}^{-2}]$\\
    $\vect{E}(\vect{x})$ & diffuse flux $[\mathrm{Wm}^{-2}]$\\
    $\mathrm{D}(\vect{x}) = \frac{1}{3\sigma_t}$ & diffusion coefficient (CDA) $[\mathrm{m}]$ \\
    $\mathrm{D_F}(\mathcal{R}, \vect{x})$ & diffusion coefficient (FLD) $[\mathrm{m}]$\\
    $\mathcal{R}(\vect{x})$ & Knudsen number \\
    $T(\vect{x})$ & transmittance/DSM \\
    $\tau$  & optical depth (number of mean free paths) \\
    \end{tabular}
\end{table}

To develop the radiative diffusion approximation, we first separate the total radiance into the sum of the reduced incident light (also referred to as the unscattered light) $L_{ri}(\mathbf{x}, \boldsymbol{\omega})$ which is  the incident radiance from external sources attenuated by extinction, and the medium radiance $L_m(\mathbf{x},\boldsymbol{\omega})$, which is generated by both scattering of the reduced incident light and emission from the medium:
\begin{equation} \label{unscatteredDiffuseDecomposition}
L(\mathbf{x}, \boldsymbol{\omega}) = L_{ri}(\mathbf{x}, \boldsymbol{\omega}) + L_m(\mathbf{x}, \boldsymbol{\omega}) \ .
\end{equation}

Substituting Eqn. (\ref{unscatteredDiffuseDecomposition}) into Eqn. (\ref{rte2}), we get a decomposition of in-scattered light, which results in:
\begin{equation} \label{source_rad_medium}
4\pi Q(\mathbf{x}) = q_{ri}(\mathbf{x}) + q_m(\mathbf{x}) + j(\mathbf{x}) \\
\end{equation}
where
\begin{equation} \label{QriQm}
\begin{split}
q_{ri}(\mathbf{x})  &\defeq \sigma_s\int_{\Omega}L_{ri}(\mathbf{x}, \omega)\ud\omega \ , \\
q_{m}(\mathbf{x}) &\defeq \sigma_s\int_{\Omega}L_{m}(\mathbf{x}, \omega)\ud\omega \ .
\end{split}
\end{equation}
The source field generated by the reduced incident light, $q_{ri}$, is determined by the choice of external lighting, and we will assume that it has been precomputed. For example in the case of an external incident directional light of radiance $L_l$ in direction $\boldsymbol{\omega_l}$, it has the simple form $q_{ri}(\mathbf{x}) = L_l \sigma_s(\mathbf{x}) T(\mathbf{x})$, where $T(\mathbf{x})$ is the transmittance from the point $\mathbf{x}$ in the volume along the ray in the direction towards the light (commonly referred to as a \emph{deep shadow map}). This can be easily generalized to other forms of external lighting.

We can think of the medium radiance $L_m(\mathbf{x}, \boldsymbol{\omega})$ as the multiply-scattered ``diffuse''  light (though it includes all orders of scattering, as well as the self-emitted radiance). The diffusion approximation consists of the assumption that we can express the medium radiance as the following expansion in the first two angular moments of the radiation field:
\begin{equation} \label{two_term_radiance_expansion}
L_m(\mathbf{x}, \boldsymbol{\omega}) \approx \frac{1}{4\pi}\phi(\mathbf{x}) + \frac{3}{4\pi}\omega\cdot \mathbf{E}(\mathbf{x}) \ .
\end{equation}
In this approximation, the diffuse light is describable entirely by two fields: the medium \emph{fluence} (or scalar irradiance, the zeroth angular moment of the medium radiance) $\phi(\mathbf{x}) = \int_{\Omega} L_m(\mathbf{x}, \mathbf{\omega}) \ud\omega$ and the medium \emph{flux} (or vector irradiance, the first angular moment of the medium radiance) $\mathbf{E}(\mathbf{x}) = \int_{\Omega} \mathbf{\omega} \, L_m(\mathbf{x}, \mathbf{\omega}) \ud\omega$. 
In fact, this \emph{is} a good approximation in highly opaque media where multiple scattering dominates.
This is because multiple scattering tends to reduce the higher-order angular moments~\cite{stam1995multiple}.  The diffusion approximation provides an efficient means of solving for $\phi(\mathbf{x})$, which determines $q_m(\mathbf{x})$ in Eqn.~\ref{QriQm}.

The source field expressed in terms of $\phi(\mathbf{x})$ is
\begin{eqnarray} \label{source_radiance_in_terms_of_phi}
4\pi Q(\mathbf{x})&=& q_{ri}(\mathbf{x}) + \sigma_s(\mathbf{x}) \phi(\mathbf{x}) + j(\mathbf{x})
\end{eqnarray}
%
\begin{figure}[!t]
\centering
\includegraphics[width=0.8\linewidth]{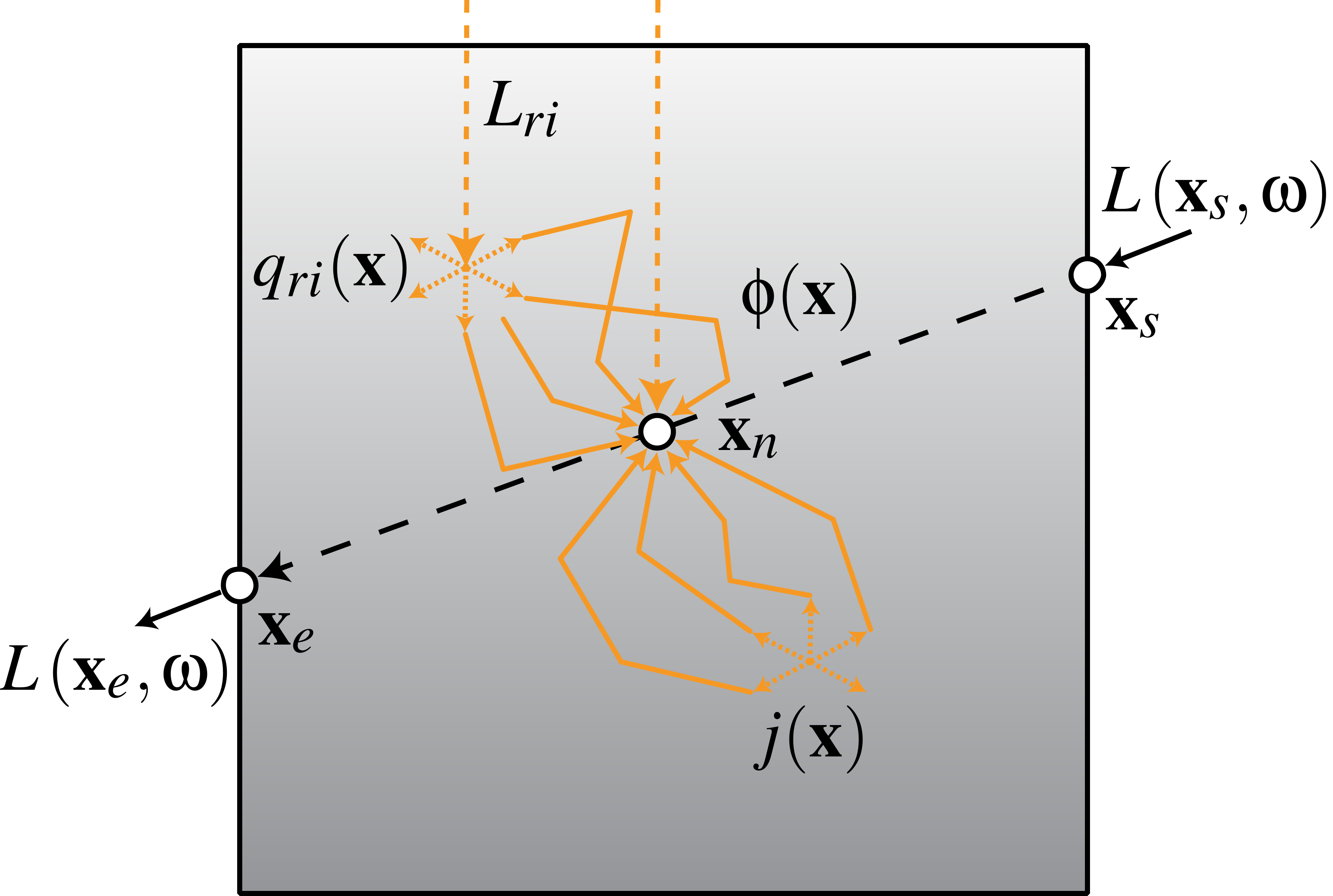}
\caption{Schematic of the volume rendering process. The unscattered attenuated incident light $L_{ri}$ generates the source field $q_{ri}(\mathbf{x})$, and emission within the volume generates the source field $j(\mathbf{x})$. The multiply-scattered light $\phi(\mathbf{x})$ is then determined by solving the diffusion Eqn.~(\ref{radiative_diffusion_equation}) with the source fields $q_{ri}(\mathbf{x})$ and $j(\mathbf{x})$. The fields $q_{ri}(\mathbf{x})$, $j(\mathbf{x})$, and $\phi(\mathbf{x})$ are then used to compute the source term $Q(\mathbf{x}_n)$ at each ray segment according to Eqn.~(\ref{source_radiance_in_terms_of_phi}), which is accumulated along the ray to compute the observed radiance according to Eqn.~(\ref{primary_raymarch}).}
\label{fig:schematic}
\end{figure}
To derive the equation for $\phi(\mathbf{x})$ in the diffusion approximation, we start with the observation that, by definition, the unscattered light satisfies the following restricted form of the RTE excluding the scattering term:
\begin{equation} \label{unscattered_rte}
(\mathbf{\omega}\cdot\nabla) L_{ri}(\mathbf{x}, \mathbf{\omega}) = -\sigma_t(\mathbf{x}) L_{ri}(\mathbf{x}, \mathbf{\omega}) \ .
\end{equation}
Thus substituting Eqn.~(\ref{unscatteredDiffuseDecomposition}) into Eqn.~(\ref{rte}),
and using Eqn.~(\ref{unscattered_rte}), we find that the medium radiance satisfies the RTE:
\begin{eqnarray} \label{diffuse_rte}
(\mathbf{\omega}\cdot\nabla) L_m(\mathbf{x}, \mathbf{\omega}) =  -\sigma_t(\mathbf{x})L_m(\mathbf{x}, \mathbf{\omega}) + Q(\mathbf{x}, \omega) \ .
\end{eqnarray}
To obtain equations for $\phi(\mathbf{x})$ and $\mathbf{E}(\mathbf{x})$, we substitute Eqn.~(\ref{two_term_radiance_expansion}) into Eqn.~(\ref{diffuse_rte}), and integrate over $\int_{\Omega}\ud\omega$ and $\int_{\Omega}\mathbf{\omega}\ud\omega$ respectively, giving:
\begin{eqnarray} \label{phi_and_E}
\nabla\cdot\mathbf{E}(\mathbf{x})  &\,=\,& -\sigma_a(\mathbf{x}) \phi(\mathbf{x}) + q_{ri}(\mathbf{x})  +  j(\mathbf{x}) \ , \nonumber \\
\mathbf{E}(\mathbf{x})   &\,=\,& -\frac{1}{3\sigma_t(\mathbf{x}) } \nabla\phi(\mathbf{x}) =  -D(\mathbf{x})\nabla\phi(\mathbf{x})\ .
\end{eqnarray}
The first line of Eqn.~(\ref{phi_and_E}) is exact and (when integrated over volume) simply states conservation of energy, i.e., that in a steady-state solution of the RTE the total power of diffuse light emitted from any given region must equal the rate of increase in diffuse radiant energy in the region due to in-scattering and emission, minus the rate of decrease due to absorption. The second line of Eqn.~(\ref{phi_and_E}), commonly termed Fick's first law of diffusion, is the statement that in the isotropic diffusion approximation the flux $\mathbf{E}$ is proportional to the gradient of $\phi$, where $D(\mathbf{x}) \defeq [3\sigma_t(\mathbf{x})]^{-1}$ is the \emph{classical diffusion coefficient}, which has dimensions of length. Flux-limited diffusion theory, to be introduced in the next section, amounts to a modification of this diffusion coefficient.

Substituting Fick's first law into the first line of Eqn.~(\ref{phi_and_E}) yields Fick's second law, the radiative diffusion equation:
\begin{eqnarray} \label{radiative_diffusion_equation}
\nabla \cdot \left( D(\mathbf{x})\nabla\phi(\mathbf{x}) \right) = \sigma_a(\mathbf{x})\phi(\mathbf{x}) - q_{ri}(\mathbf{x}) -  j(\mathbf{x}) \ .
\end{eqnarray}
This describes how the zeroth moment of the multiply-scattered radiance (or fluence), $\phi(\mathbf{x})$, is determined by the incident unscattered light and the emission, represented by the source terms $q_{ri}(\mathbf{x})$ and $j(\mathbf{x})$. 
This is a partial differential equation defining a boundary value problem which can be solved for $\phi(\mathbf{x})$ in the volume domain, given specified conditions on $\phi(\mathbf{x})$ to impose at the domain boundary. We use Dirichlet boundary conditions where $\phi(\mathbf{x})$ is simply set to zero on the boundary. Though this is physically inaccurate, our procedure of using the unscattered light to seed the diffusion solve in the interior of the volume minimizes the influence of the boundaries. 

Volume rendering an image  is then completed by marching through the volume along the primary ray in order to compute the sum in Eqn.~(\ref{primary_raymarch}), using the computed fluence field $\phi(\mathbf{x})$. The whole process is summarized in Fig.~\ref{fig:schematic}.

\section{Flux-Limited Diffusion}
\label{sec:fld}

In this section, we introduce the theoretical ideas of flux-limited diffusion theory (FLD), an enhancement to the classical diffusion theory which proves to be more accurate, especially in heterogeneous media with low opacity regions. In outline, the method amounts to replacing the diffusion coefficient $D(\mathbf{x}) = [3\sigma_t(\mathbf{x})]^{-1}$ with an expression which depends on the fluence $\phi(\mathbf{x})$ and the magnitude of its gradient, $|\nabla\phi(\mathbf{x})|$. FLD was developed originally for radiative transfer applications in astrophysics \cite{Levermore_Pomraning_1981, 2001ApJS..135...95T} and nuclear physics \cite{1974JMP....15...75L}. Here, we will state the mathematical framework of FLD required for its application, but without giving the full mathematical derivation, aiming instead to give intuition for its physical meaning.

The classical diffusion approximation cannot be applied straightforwardly to the computation of multiple scattering in 
heterogeneous media, as it breaks down and violates the true RTE in regions of low extinction $\sigma_t$ (or equivalently large mean free path).
The first indication of this is that in the limiting case of a vacuum (i.e., infinite mean-free-path), the radiative diffusion equation, Eqn.~(\ref{radiative_diffusion_equation}), becomes singular, because the diffusion coefficient $D(\mathbf{x})$ diverges in the limit $\sigma_t\rightarrow 0$. 
A more rigorous demonstration that this is inconsistent with the true RTE is as follows. The medium radiance is non-negative and therefore manifestly satisfies the following inequality, where $\mathbf{\hat{n}}$ is any unit vector:
\begin{equation}
\int_{\Omega} \left( 1 - \mathbf{\hat{n}}\cdot\mathbf{\omega}  \right)  L_m(\mathbf{x}, \boldsymbol{\omega})\ud\omega\ge 0 \ .
\end{equation}
This reduces to the statement that $\phi(\mathbf{x}) \ge \mathbf{\hat{n}}\cdot\mathbf{E}(\mathbf{x})$ for all unit vectors $\mathbf{\hat{n}}$, which implies the constraint
\begin{equation} \label{eq:flux_limiting_constraint}
\left | \mathbf E(\mathbf x)\right | \le \phi(\mathbf x) \ .
\end{equation}
This states that at any point in the radiation field the magnitude of the flux does not exceed the fluence. Physically this is related to the fact that radiation propagates at the finite speed of light, so energy cannot be transported at an arbitrarily high rate from point to point (thus this flux constraint is sometimes said to enforce ``causality'').

In CDA, $\phi(\mathbf{x})$ satisfies Eqn.~(\ref{radiative_diffusion_equation}) and the corresponding flux is given by the Fick's law relationship (the second line of Eqn.~(\ref{phi_and_E})). However, there is no mechanism to prevent a solution where $\left | \mathbf E(\mathbf{x})\right |=|D(\mathbf{x})\nabla\phi(\mathbf{x})|$   is arbitrarily large compared to $\phi(\mathbf{x})$, which violates the constraint in Eqn.~(\ref{eq:flux_limiting_constraint}). This suggests that CDA can be improved by explicitly enforcing the  constraint, which is the core idea of FLD theory.

To achieve this, in FLD theory the diffusion coefficient in Fick's law is modified to
\begin{equation} \label{fld_diffusion_coeff}
D_F(\mathbf{x}) = \frac{\mathcal{F(\mathcal{R}(\mathbf{x}))}}{\sigma_t(\mathbf{x})} \ ,
\end{equation}
where the dimensionless $\mathcal{F}(\cdot)$ is termed the \emph{flux limiter}. CDA corresponds to setting $\mathcal{F}$ to the constant $1/3$. In FLD, $\mathcal{F}$ is a function of the dimensionless variable $\mathcal{R}(\mathbf{x})$, termed the  \emph{Knudsen number}, which is defined as follows:
\begin{equation}
\label{eq:fld_R}
\mathcal{R}(\mathbf x) \defeq \frac{\left|\nabla\phi(\mathbf x)\right|}{\sigma_t(\mathbf{x})\phi(\mathbf x)} \ .
\end{equation}
Via the modified Fick's first law $\mathbf{E}(\mathbf{x})  =  -D_F(\mathbf{x})\nabla\phi(\mathbf{x})$, the flux-limiting constraint Eqn.~(\ref{eq:flux_limiting_constraint}) expressed in terms of the flux limiter and Knudsen number becomes:
\begin{equation} \label{eq:F_constraint}
\mathcal{F} \le \frac{1}{\mathcal{R}} \ .
\end{equation}

The physical meaning of the Knudsen number $\mathcal{R}$ can be understood roughly as:
\begin{equation}
\label{eq:fld_R_intuitive}
\mathcal{R} \sim \frac{\mbox{mean free path}}{\mbox{length scale of $\phi$ variations}} \ .
\end{equation}
When the local mean free path is short relative to the local length scale of changes in the fluence, $\mathcal{R}\ll 1$ and the radiation field is in a ``diffusive'' regime where CDA, $\mathcal{F}=1/3$, is a good approximation. In this case Eqn.~(\ref{eq:F_constraint}) is satisfied automatically.
However, when the local mean free path is much longer than this length scale, $\mathcal{R}\gg 1$ and the radiation field enters the ``transport'' regime where interaction events are rare over the length scale being considered. Thus in FLD, in order to satisfy the constraint Eqn.~(\ref{eq:F_constraint}), we use a modified flux limiter which satisfies the limits:
\begin{equation}
\label{eq:F_limits}
\lim_{\mathcal{R}\rightarrow 0}      \;\mathcal{F}(\mathcal{R}) \; = \frac{1}{3} \ , \quad
\lim_{\mathcal{R}\rightarrow \infty} \mathcal{R}\;\mathcal{F}(\mathcal{R}) \; = 1 \ .
\end{equation}

A variety of methods have been developed to find specific functional forms for $\mathcal{F}(\mathcal{R})$ satisfying the constraints in Eqn.~({\ref{eq:F_limits}). 
Some approaches use forms which simply satisfy the desired limits \cite{1974JMP....15...75L, 1982ApJS...50..115B}. Others derived flux limiters based on certain assumptions about the smoothness of the moments \cite{Levermore_Pomraning_1981, Kershaw76}. We give the expressions for these simple widely used flux limiters in Table~\ref{tab:flux_limiters}.
Other strongly related work includes Minerbo's \cite{1978JQSRT..20..541M} flux limiters which are based on statistical mechanics of the radiation field, and a method termed ``M1-theory'' which works by making assumptions about the local rotational symmetry of the radiation field and deriving the corresponding pressure tensor closure \cite{2000JQSRT..64..619O}. We do not expand on either of these as this would exceed the scope of this paper.

\begin{table}[!t]
\caption{Flux limiters, $\mathcal{F}(\mathcal{R})$  \label{tab:flux_limiters}.}
\vspace{-0.3cm}
\centering
      \begin{tabular}{ll} 
    \\ \hline\hline
    Flux limiter                                                   & $\mathcal{F}(\mathcal{R})$ 					
   	\\ \hline
        sum \cite{1982ApJS...50..115B}           &  $(3+\mathcal{R})^{-1}$                                         \\
        max   \cite{1982ApJS...50..115B}         &  $\mbox{max}(3, \mathcal{R})^{-1}$                       \\
        Kershaw \cite{Kershaw76}                    & $2(3+\sqrt{9 + 4\mathcal{R}^2}\,)^{-1}$      \\
        Larsen-$n$ \cite{1974JMP....15...75L} & $(3^n + \mathcal{R}^n)^{-\frac{1}{n}}$                    \\
        Levermore--Pomraning \cite{Levermore_Pomraning_1981} &  $\frac{1}{\mathcal{R}} \left(\coth(\mathcal{R})-\frac{1}{\mathcal{R}}\right)$  \\ 
	 \hline
    \end{tabular}
\end{table}

Regardless of the specific choice of flux limiter, the resulting flux-limited radiative diffusion equation has the form
\begin{eqnarray} \label{fld_radiative_diffusion_equation}
\nabla \cdot \left( D_F(\mathbf{x})\nabla\phi(\mathbf{x})\right) = \sigma_a(\mathbf{x})\phi(\mathbf{x}) - q_{ri}(\mathbf{x})  -  j(\mathbf{x}) \ .
\end{eqnarray}
Here $D_F$ implicitly depends on $\phi(\mathbf{x})$ via Eqn.~(\ref{eq:fld_R}), which makes this a non-linear PDE, unlike the classical diffusion Eqn.~(\ref{radiative_diffusion_equation}) which is linear. As in the CDA case, we use Dirichlet boundary conditions to specify $\phi(\mathbf{x})$ on the domain edge. In summary, FLD takes the theoretical framework of CDA and modifies it straightforwardly by replacing the diffusion coefficient with a $\phi(\mathbf{x})$-dependent expression. In Sec.~\ref{sec:numerical_method}, we give the details of implementing a numerical CDA solver and extending it to support the FLD technique. 

Note that in the limit $\mathcal{R}\rightarrow \infty$, the divergence term in the diffusion equation becomes $-\nabla\cdot(\phi(\mathbf{x})\hat{\mathbf{n}})$, which effectively transforms the diffusion equation into an advection equation for the fluence. 
This also explains why the flux limiter is chosen to  tend to exactly $1/\mathcal{R}$ in Eqn.~(\ref{eq:F_limits}), which is a stronger requirement than imposed by the constraint of Eqn.~(\ref{eq:F_constraint}): FLD  not only suppresses the flux in the transport regime, it also saturates it at the appropriate value to ensure correct free propagation (at the level of the approximation).

\begin{figure*}[!ht]
  \centerline{
  \hfill
\subfigure[Grosjean point source solution\label{subfig:analytical_fld}]{\includegraphics[width=0.327\linewidth]{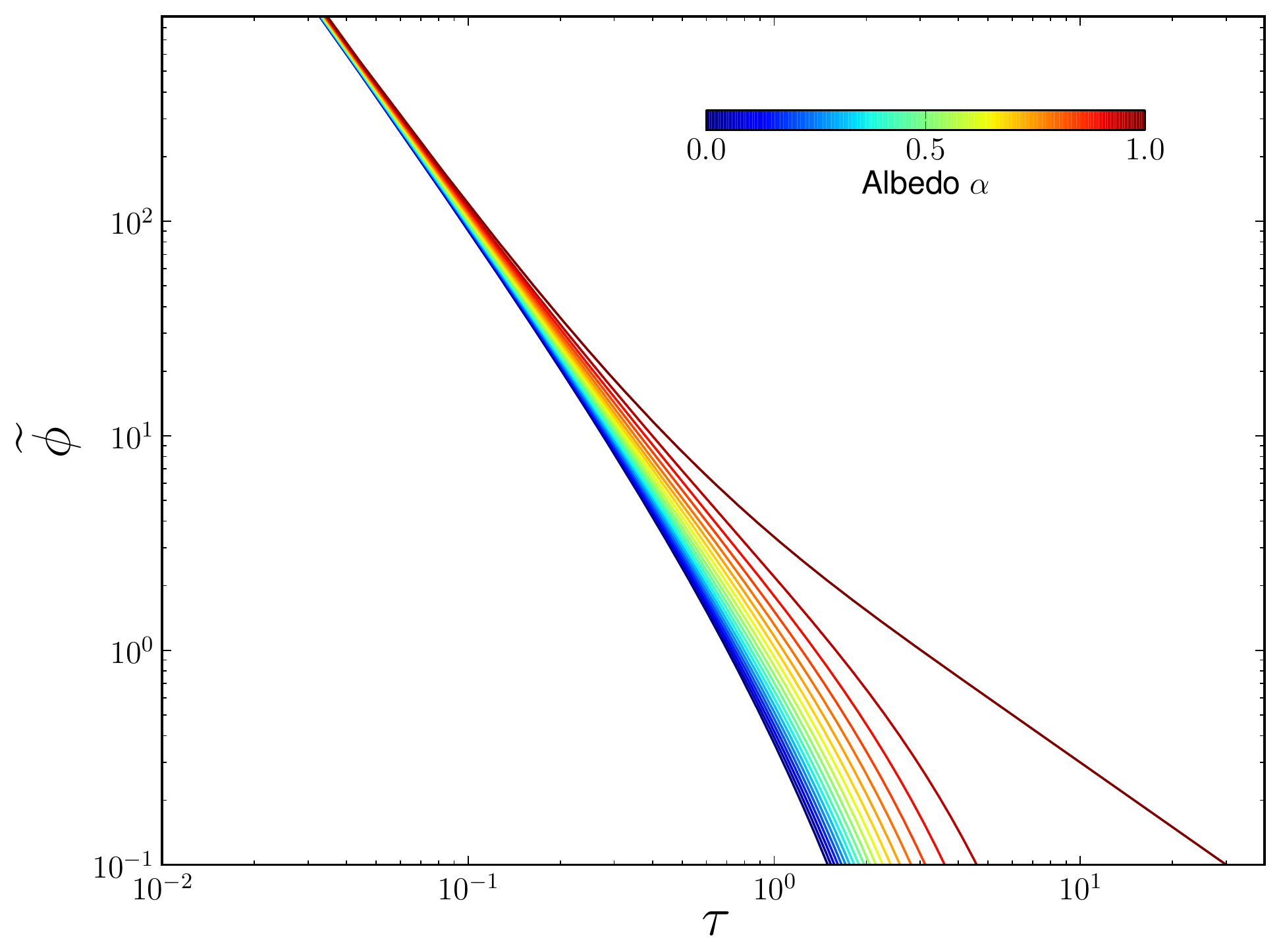}\label{subfig:grosjean_analytical}}
  \hfill
\subfigure[CDA point source solution\label{subfig:analytical_cda}]{\includegraphics[width=0.327\linewidth]{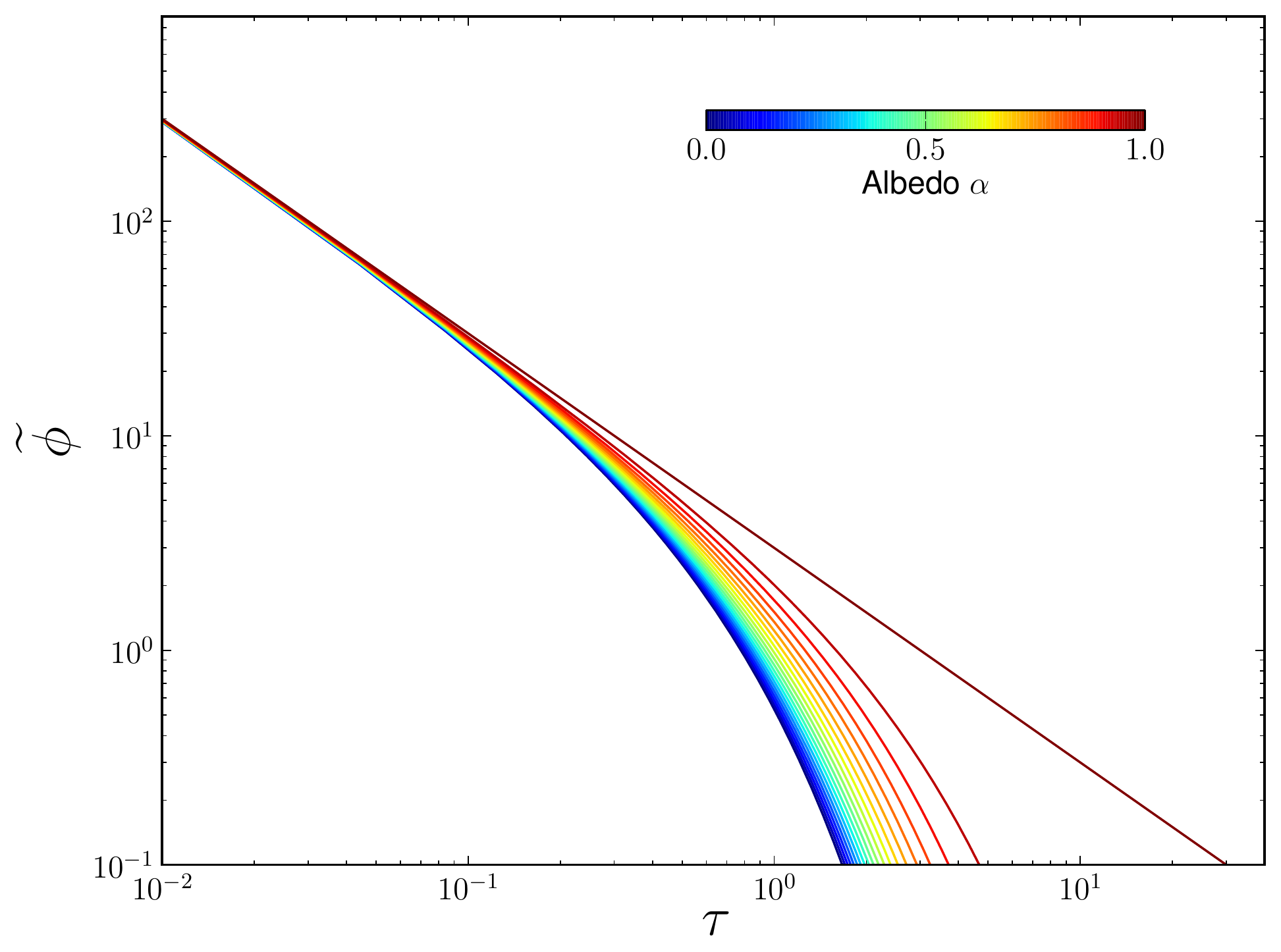} \label{subfig:cda_analytical}}
  \hfill
\subfigure[Knudsen number, $\mathcal{R}$\label{subfig:analytical_R}]{\includegraphics[width=0.327\linewidth]{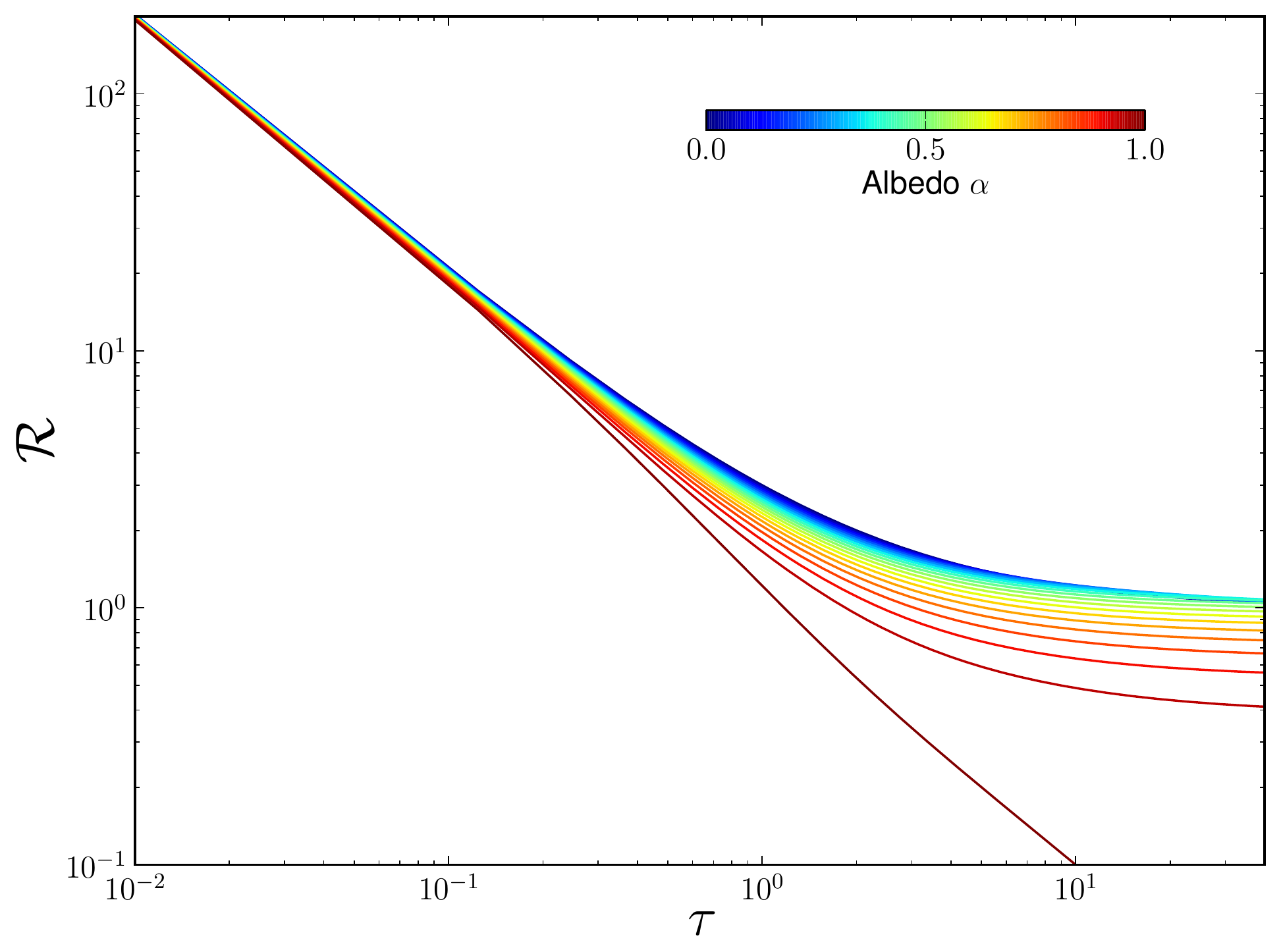}\label{subfig:knudsen_analytical}}
\hfill}
\caption{Analytical ``universal'' solutions for normalized fluence versus radius for a unit power point source embedded in a homogeneous medium. We plot in \subref{subfig:analytical_fld} the Grosjean transport solution Eqn.~(\ref{eq:grosjean}) and in \subref{subfig:analytical_cda} the classical diffusion approximation Eqn.~(\ref{eq:cda_greensfunction}), both for a range of albedos between $0$ and $1$. To show the solutions for all possible $\sigma_t$ we plot normalized fluence (defined in Eqn.~(\ref{eq:normalized_grosjean_fluence})) as a function of optical depth $\tau=\sigma_t r$. In \subref{subfig:analytical_R} we plot the Knudsen number $\mathcal{R}$ (Eqn.~(\ref{eq:fld_R})) of the Grosjean solution.}
\label{fig:point_source_theory}
\end{figure*}

A deeper examination of the derivation of CDA gives further insight into the physical content of FLD.
As discussed in Section \ref{sec:cda},  CDA can be derived by expressing the radiance in terms of its first two moments only, leading to the diffusion equation which determines these two moments. It was not specified what assumptions we made about the higher moments---in general these moments are non-zero as they are coupled to the first and second moments via a hierarchy. In fact, a deeper description of the diffusion approximation is that the moment hierarchy is closed by choosing a specific form for the second moment, termed the \emph{radiation pressure tensor}, $P_{ij}(\mathbf{x}) = \int_{\Omega} \omega_i \omega_j \, L_m(\mathbf{x}, \mathbf{\omega}) \ud\omega$. In the case of CDA, the radiation pressure tensor is approximated as isotropic, $P_{ij} = \delta_{ij}\phi/3$, which leads to the approximate two-term expansion of the radiance given in Eqn.~(\ref{two_term_radiance_expansion}). 

In the case of FLD, the form of the pressure tensor is different. In terms of the \emph{Eddington tensor} $E_{ij}$, defined by $P_{ij} = E_{ij} \phi$, in CDA $E_{ij} = \delta_{ij}/3$, while in FLD
\begin{equation}
E_{ij} = \frac{1}{2}(1 - \mathcal{X}) \delta_{ij} + \frac{1}{2}(3\mathcal{X}-1) \, \mathbf{\hat{n}}_i \mathbf{\hat{n}}_j \ ,
\end{equation}
where $\mathbf{\hat{n}}$ is the
unit vector in the direction of the negative fluence gradient, $\mathbf{\hat{n}}(\mathbf{x}) = -\nabla\phi(\mathbf{x})/|\nabla \phi(\mathbf{x})|$, and $\mathcal{X}$ is a dimensionless function called the \emph{Eddington factor}. In a full analysis it has been shown \cite{Pomraning1982517, 1984JQSRT..31..149L} that the following approximate relationship 
holds between the flux limiter $\mathcal{F}$ and the Eddington factor $\mathcal{X}$:
\begin{equation}
\mathcal{X}= \mathcal{F} + \mathcal{F}^2 \mathcal{R}^2 \ .
\end{equation}
This implies that in the diffusive and transport limits respectively, the FLD Eddington tensor becomes
\begin{equation}
\label{eq:D_F_limits}
\lim_{\mathcal{R}\rightarrow 0}     \;E_{ij} = \;\delta_{ij}/3 \ , \quad
\lim_{\mathcal{R}\rightarrow \infty} \;E_{ij}  = \; \mathbf{\hat{n}}_i \mathbf{\hat{n}}_j \ .
\end{equation}
From this we see explicitly that in the transport regime, FLD produces a radiation pressure tensor closure which corresponds to free propagation of rays down the local energy density gradient.  
It should be borne in mind however that though FLD deals in a more physically sensible manner 
with the transition to the transport regime than CDA, it is still just an approximation to the full RTE, as all but the first few moments of the radiation field are used. 

\subsection{Point Source in Homogeneous Medium} \label{sec:point_source_theory}

In this section we apply the ideas introduced above to the problem of a unit power point light embedded in an infinite homogeneous scattering medium with extinction coefficient $\sigma_t$ and albedo $\alpha$. This is a spherically symmetric problem for which the exact solution is known to both the full RTE and CDA. We use this example both to illustrate the general idea of FLD and as a testbed for our numerical solver.

The exact solution of the RTE for the fluence as a function of radial distance from the point source, and an excellent approximation to it, termed the Grosjean solution~\cite{grosjean1959multiple}, are quoted in previous work~\cite{d2011quantized}. 
Parameterizing the radial distance $r$ from the point source in terms of the dimensionless \emph{optical depth}  $\tau = \sigma_t r$, the Grosjean solution for the fluence is given (in the case of an isotropic phase function) by:
\begin{equation}  \label{eq:grosjean}
\phi\left(\tau\right) = \frac{\sigma_t^2}{4\pi} \left( \frac{\euler^{-\tau}}{\tau^2} + \frac{3\alpha}{2-\alpha}\frac{\euler^{-\lambda \tau}}{\tau} \right) \ ,
\end{equation}
where $\lambda^2 \defeq \frac{3(1-\alpha)}{2-\alpha}$. Note that the $\sigma_t^2$ factor occurs since $\phi$ has units of radiance (power per unit area), and we used a unit power source. The expression inside the brackets is dimensionless. In Fig.~\ref{subfig:grosjean_analytical}, we plot the ``normalized fluence''
\begin{equation} \label{eq:normalized_grosjean_fluence}
\widetilde{\phi}(\mathbf{\tau}) \defeq 4\pi \phi(\tau)/\sigma_t^2 \ .
\end{equation}
as a function of $\tau$. These figures show the $\sigma_t$-independent ``universal'' solution to the point source scattering problem.

The CDA in Eqn.~(\ref{radiative_diffusion_equation}) reduces in this homogeneous emission-only case to a screened Poisson equation:
\begin{equation}
\label{eq:da}
D \nabla^2\phi(\mathbf{x}) = \sigma_a(\mathbf{x})\phi(\mathbf{x}) - j(\mathbf{x})  \ .
\end{equation}
The solution to this, for a unit power point source emission $j(\mathbf{x}) = \delta(\mathbf{x})$, known as the CDA Greens function~\cite{jensen2001practical}, is given by
\begin{equation} \label{eq:cda_greensfunction}
\phi\left (\tau\right) = \frac{3\sigma_t^2}{4\pi}\frac{\euler^{-\sqrt{3\left ( 1-\alpha\right )}\tau}}{\tau} \ .
\end{equation}
Fig.~\ref{subfig:cda_analytical} plots the CDA Greens function as a function of optical depth. As it has been discussed in detail~\cite{d2011quantized}, we see that CDA fails to generate the $\tau^{-2}$ power law term in the Grosjean solution, and also has stronger exponential suppression of the $\tau^{-1}$ diffusion power law at large radii, which causes it to underestimate the fluence. 

In the Grosjean solution the Knudsen number $\mathcal{R}\rightarrow 2/\tau$ as $\tau\rightarrow 0$ (see Fig.~\ref{subfig:knudsen_analytical}). Thus, sufficiently close to the point source, i.e., roughly within a mean-free-path, the radiation field is in the transport regime according to the criterion described in Section~\ref{sec:fld}. 
In this region, the inverse square term in Eqn.~(\ref{eq:grosjean}) dominates (and in the $\alpha=0$ limit it is the exact transport solution, as can be verified by trivial application of the RTE without scattering). 
In the terminology of~\cite{d2011quantized}, this term represents the \emph{ballistic fluence} due to light propagating unscattered from the source. 

As pointed out by Levermore and Pomraning~\cite{Levermore_Pomraning_1981}, in spherical symmetry the FLD Eqn.~(\ref{fld_radiative_diffusion_equation}) with a point emission source becomes in the transport regime (according to the $\mathcal{R}\rightarrow\infty$ limiting form in Eqn.~(\ref{eq:F_limits})) the following linear first-order ordinary differential equation:
\begin{equation}
\label{eq:fld3}
\frac{1}{\tau^2}\frac{\mathrm{d}}{\mathrm{d}\tau}(\tau^2\phi(\tau)) + (1-\alpha)\phi(\tau) = j(r)/\sigma_t \ ,
\end{equation}
whose corresponding solution for a delta function $j$ is
\begin{equation}
\phi(\tau) = \frac{\sigma_t^2}{4\pi}  \frac{\euler^{-(1-\alpha)\tau}}{\tau^2} \ ,
\end{equation}
which tends to the ballistic term in the Grosjean solution as $\tau\rightarrow 0$. This verifies that FLD correctly reproduces the transport limit for the point source problem. Thus, one reasonable test of our formulation is to check that FLD gives this inverse square behavior of the fluence near an emitting point source. However, note that we do not expect FLD to perfectly match the transport theory throughout the domain, as the FLD equation is merely an approximation to the RTE.

\section{Numerical Method}
\label{sec:numerical_method}

We provide here a numerical method for solving Eqn.~(\ref{fld_radiative_diffusion_equation}) to obtain the multiply-scattered fluence $\phi(\mathbf{x})$. We assume a uniform 3D grid with voxel centers indexed by integer coordinates $p=(i,j,k)$, and voxel edge length $\Delta l$. For notational convenience, we absorb the external illumination term $q_{ri}(\mathbf{x})$ into the emission term $j(\mathbf{x})$. Hence, the voxelized input fields are the emission field $j_p$ (units of power per unit volume), the extinction field $\sigma_p\defeq\left.\sigma_t\right|_{ijk}$ (units of inverse length), and the albedo field $\albedo_p$ (dimensionless). For brevity we write the flux-limited diffusion coefficient $\left.D_F\right|_{ijk}$ as $D_p$ (units of length).

Discretizing the LHS of Eqn.~(\ref{fld_radiative_diffusion_equation}) at each grid point, the $x$-derivative in the divergence term reads
\begin{equation}
\frac{D_{i+1/2, j,k} (\phi_{i+1,j,k}-\phi_p) - D_{i-1/2, j,k} (\phi_p-\phi_{i-1,j,k})}{\Delta l^2} \ ,
\end{equation}
where $D_{i+1/2, j,k}$ denotes the diffusion coefficient interpolated linearly at $(i+1/2,j,k)$. Thus, defining $s(p)$ as the 6-point stencil surrounding point $p$, so $s(p)$ ranges over the six points $(i\pm 1,j\pm 1,k\pm 1)$, we obtain the following discretized form of Eqn.~(\ref{fld_radiative_diffusion_equation}):
\begin{equation}  \label{discretized_fld_diffusion_eqn}
\frac{\sideset{}{_s}{\sum} D_{ps} \phi_s - \phi_p \sideset{}{_s}{\sum}  D_{ps}}{\Delta l^2} = (1-\alpha_p)\sigma_p \phi_p - j_p \ ,
\end{equation}
where we abbreviated $D_{ps} \defeq (D_p + D_s)/2$. Given the current set of $\phi_p$ values on the grid, we can estimate the accuracy of the solution by the \emph{residual} $R_p$, which is defined as the LHS minus the RHS of Eqn.~(\ref{discretized_fld_diffusion_eqn}). Our goal is to solve for a set of values of $\phi_p$ at every interior grid point which minimizes the root mean square (RMS) residual. However, in order to apply the FLD method, $D_p$ at each point must also satisfy Eqn.~(\ref{fld_diffusion_coeff}), which makes Eqn.~(\ref{discretized_fld_diffusion_eqn}) a non-linear equation involving the values of $\phi$ and its partial derivatives on the stencil (the specific form of this equation depends on the choice of the flux limiter $\mathcal{F}$).

\begin{algorithm}[!b] \label{redblack_sor_relaxation_algo}
\DontPrintSemicolon
\SetKwFor{parallelfor}{parallelfor}{do}{end}
Input voxel grids $j_p$, $\sigma_p$, $\alpha_p$ of extent $L$, voxel size $\Delta l$\;
Impose extinction tolerance $\sigma_p \leftarrow \mbox{max}(\sigma_p, \sigma_\epsilon)$ \;
Compute RMS emission field $\bar{j}$ \;
Initialize voxel grids $\phi_p = \epsilon \bar{j}\Delta l$, $D_p = \epsilon \Delta l$, $\forall p$ \;
Choose flux limiter $\mathcal{F}(\mathcal{R})$, from Table~\ref{tab:flux_limiters} \;
Choose SOR factor, $0<\omega<2$ \;
\Repeat{ $\bar{R}$ < $10^{-6} \bar{j}$ }
{
	\parallelfor{$\forall p$ (non-boundary red points)}
	{
		Compute $\mathcal{R}$ via Eqn.~(\ref{knudsen_numerical}) \;
		Update $D_p$ via Eqn.~(\ref{D_update}), using $\mathcal{F}(\mathcal{R})$ \;  
		Compute $D_{ps} = (D_p + D_s)/2$ $\forall$ 6 points $s$ \;
		Compute $\phi'_p$ via Eqn.~(\ref{phi_update}), using $D_{ps}$ \; 
		Update $\phi_p \leftarrow \omega \phi'_p + (1-\omega)\phi_p$,  Eqn.~(\ref{sor_phi_update}) \;
		Compute $R_p$  via Eqn.~(\ref{R_calc}) \;  
	}

	\parallelfor{$\forall p$ (non-boundary black points)}
	{
		As above.
	}
	
	Compute RMS of residual field, $\bar{R}$ \;
}
\caption{Gauss-Seidel FLD solver}
\end{algorithm}

\begin{figure*}[!tbh]
  \centerline{
  \hfill
\subfigure[\label{subfig:convergence_sor}]{\includegraphics[width=0.32\linewidth]{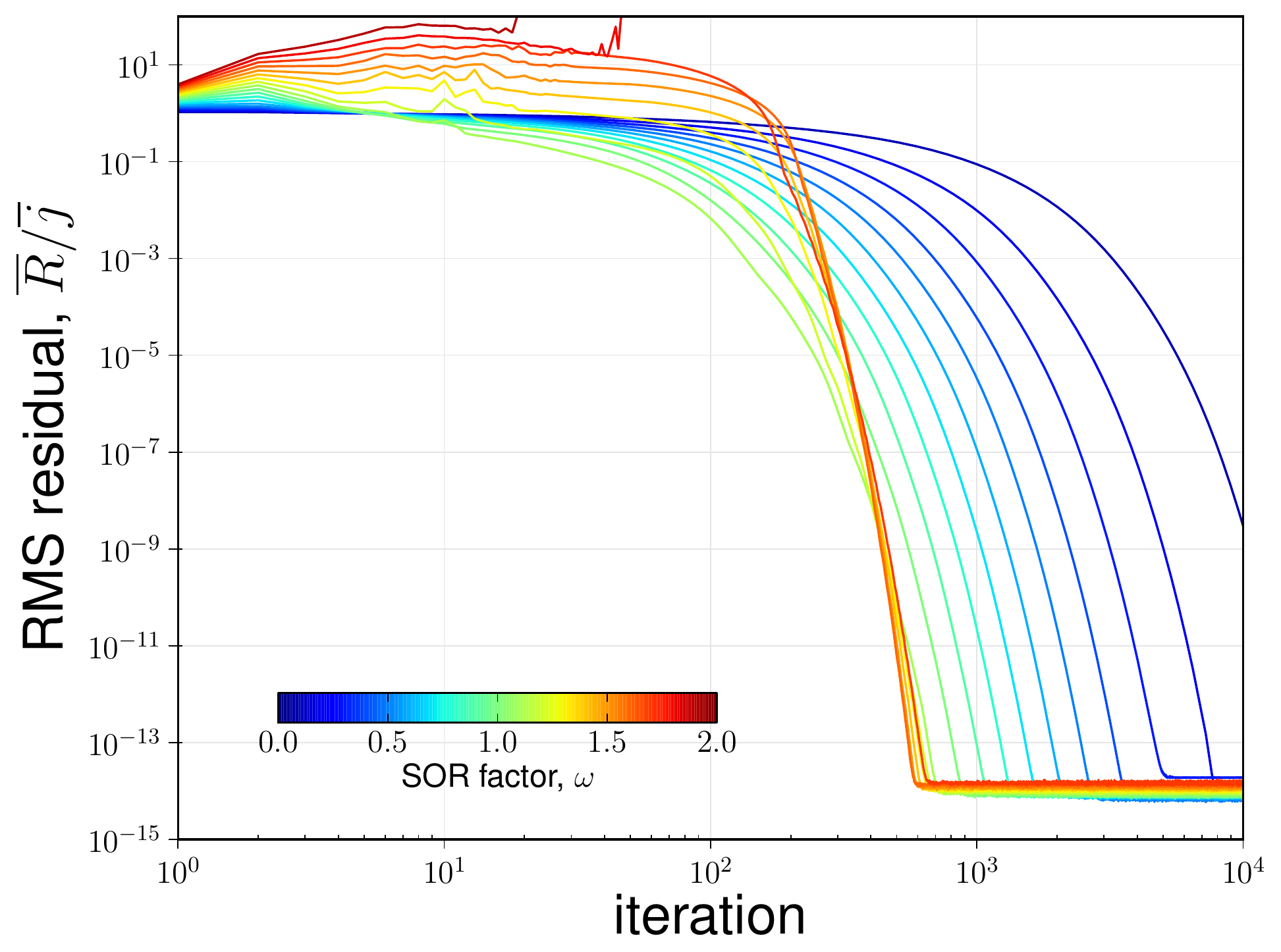}\label{subfig:residual-against_sor}}
  \hfill
\subfigure[\label{subfig:convergence_gridsize}]{\includegraphics[width=0.32\linewidth]{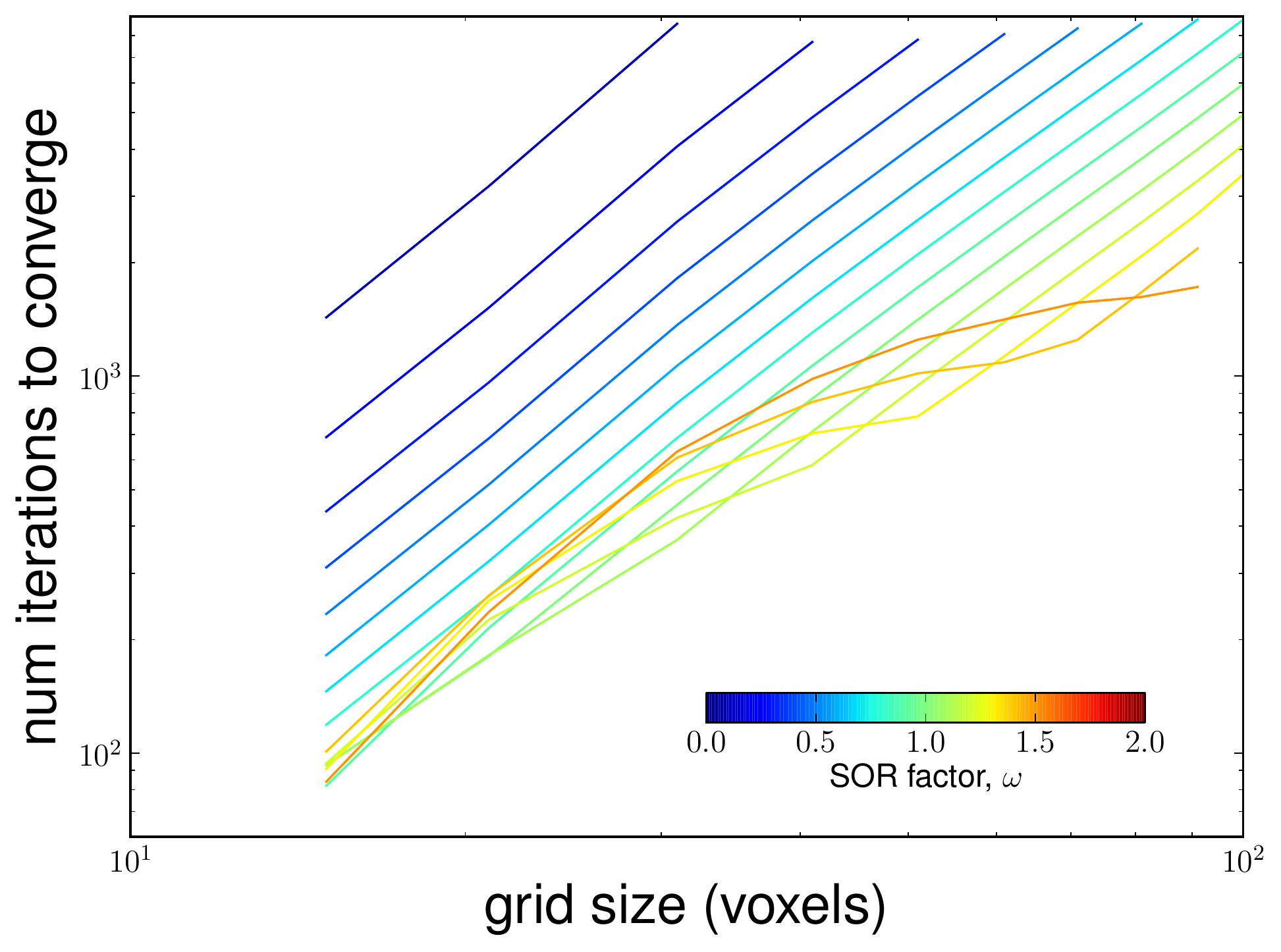}\label{subfig:residual-against_iterations}}
  \hfill
\subfigure[\label{subfig:convergence_residual}]{\includegraphics[width=0.32\linewidth]{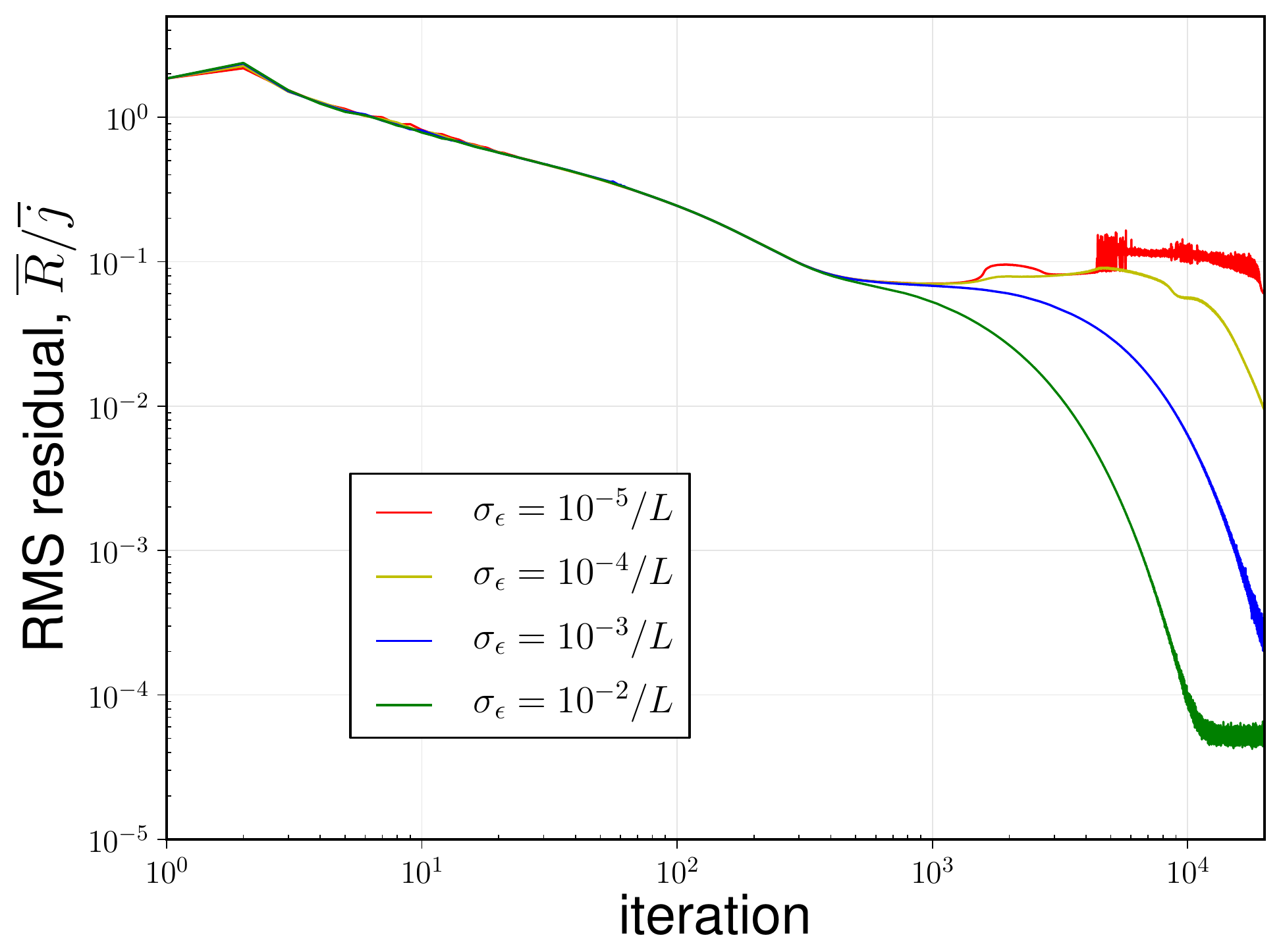}\label{subfig:kt_min_wedge}}
\hfill
}  
\caption{Convergence behavior of our FLD solver. In \subref{subfig:convergence_sor} we show the normalized RMS residual $\bar{R}/\bar{j}$ as a function of iterations for the problem described in Section~\ref{sec:numerical_method}, run in double precision, with a $51^3$ grid. In \subref{subfig:convergence_gridsize} we show the number of iterations required to reach the $\bar{R}/\bar{j}=10^{-6}$ convergence criterion as a function of grid resolution for the same test case. 
In \subref{subfig:convergence_residual} we demonstrate the effect of extinction tolerance $\sigma_\epsilon$ on convergence of the Nebulae dataset in Fig.~\ref{fig:nebulae}, here in single precision.}
  \label{fig:fld_solver_convergence}
\end{figure*}

To find a consistent solution for both $\phi_p$ and $D_p$, we employ a Gauss-Seidel relaxation scheme. We solve Eqn.~(\ref{discretized_fld_diffusion_eqn}) for $\phi_p$ to obtain the following local update rule for $\phi_p$:
\begin{equation}  \label{phi_update}
\phi_p \leftarrow \frac{j_p \Delta l^2 + \sideset{}{_s}{\sum} D_{ps} \phi_s } {(1-\alpha_p)\sigma_p \Delta l^2 + \sideset{}{_s}{\sum}  D_{ps}} \ .
\end{equation}
We also update $D_p$ at each stencil according to
\begin{equation}  \label{D_update}
D_p \leftarrow \frac{\mathcal{F}(\mathcal{R}_p)}{\mbox{max}(\sigma_p, \sigma_\epsilon)} \ .
\end{equation}
The threshold $\sigma_\epsilon$ introduced here is discussed later in this section. The local Knudsen number $\mathcal{R}_p$ is computed using the gradient vector at $p$, which is discretized as
\begin{equation}  \label{grad_approx}
\nabla\phi_p = \frac{1}{2\Delta l}\left(
\begin{array}{c}
\phi_{i+1,j,k}-\phi_{i-1,j,k} \\
\phi_{i,j+1,k}-\phi_{i,j-1,k} \\
\phi_{i,j,k+1}-\phi_{i,j,k-1} \\
\end{array}
\right) \ .
\end{equation}
We then compute
\begin{equation} \label{knudsen_numerical}
\mathcal{R}_p = \frac{\mbox{max}(\left|\nabla\phi_p\right|, \epsilon\bar{j})}{\mbox{max}(\sigma_p\phi_p, \epsilon\bar{j})} \ .
\end{equation}
This sets a lower bound $\epsilon\bar{j}$ on $\left|\nabla\phi_p\right|$ and $\sigma_p\phi_p$, where $\epsilon=10^{-20}$ and $\bar{j}$ is the RMS value of the emission field, to deal with the case when either $\phi_p$ or $\left|\nabla\phi_p\right|$ is extremely small.

For parallelization we use a Red-Black Gauss-Seidel update process. We color the grid points as a red-black checkerboard, then apply the updates of Eqn.~(\ref{phi_update}) and~(\ref{D_update}) concurrently at all red points (one ``pass'') followed by a pass over all black points. Each stencil computes and updates only its own local $\phi_p$ and $D_p$ values during the red (black) pass, which is then stored in the grid for use in the following black (red) pass. The two passes constitute one iteration. The $\phi_p$ and $D_p$ grids are initialized to small tolerances $\phi_p = \epsilon \bar{j}\Delta l$ and $D_p = \epsilon \Delta l$. Initially, these grid values do not satisfy Eqn.~(\ref{phi_update}) and Eqn.~(\ref{D_update}), but over a number of iterations they converge to a consistent solution of both equations over the whole grid. 
 At the grid boundaries, we impose Dirichlet boundary conditions with $\phi=\epsilon \bar{j}\Delta l$ by ignoring the voxels abutting the six grid faces during the update. This introduces some unphysical effects near the boundaries, but does not significantly corrupt the solution in the bulk of the grid.

\begin{figure*}[!tbh]
\hspace*{\fill}
\subfigure[CDA (exact boundary conditions)\label{subfig:numerical_cda2}]{\includegraphics[width=0.32\linewidth]{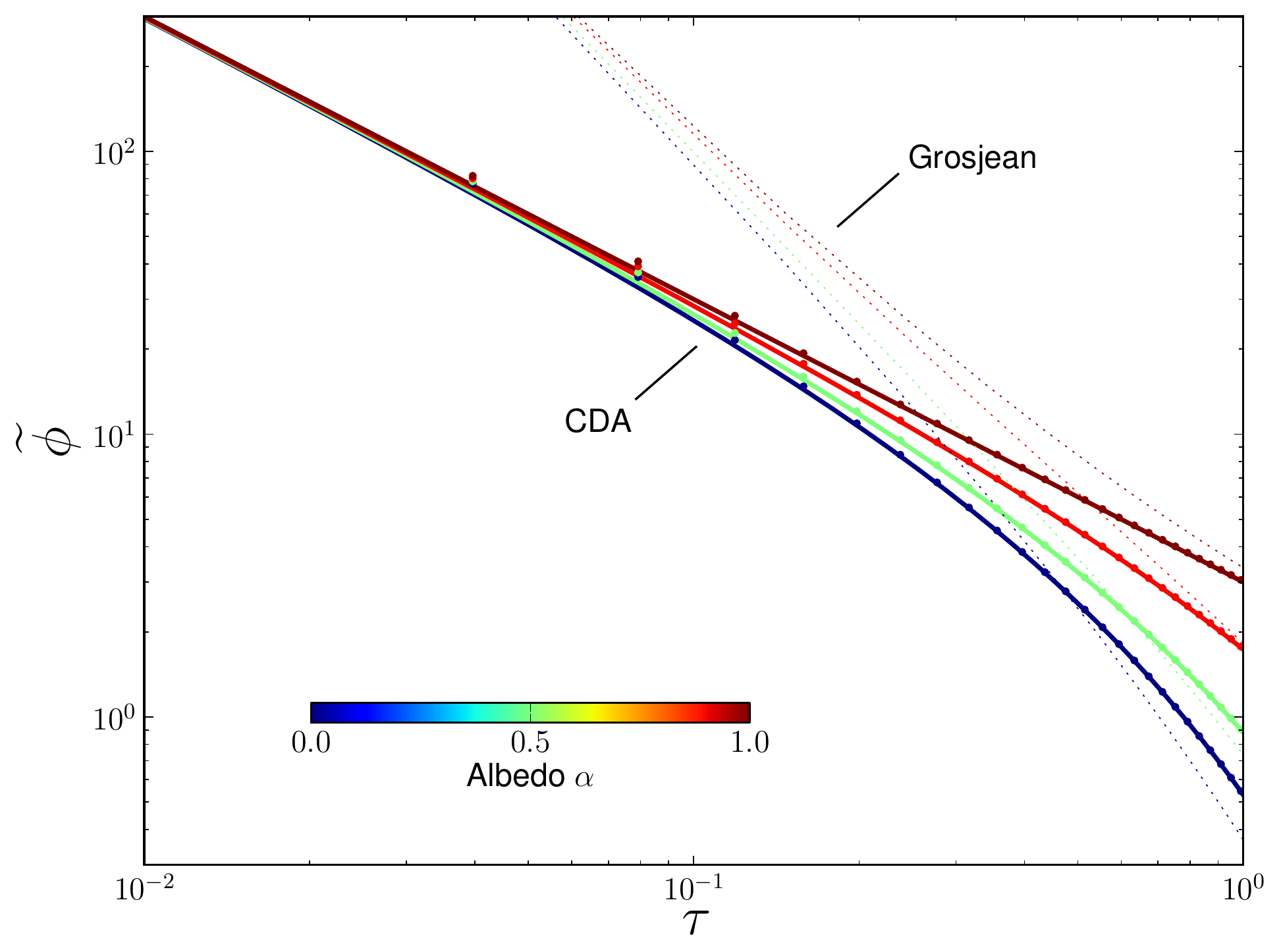} \label{CDA dirichletB}}
\subfigure[CDA\label{subfig:numerical_cda}]{\includegraphics[width=0.32\linewidth]{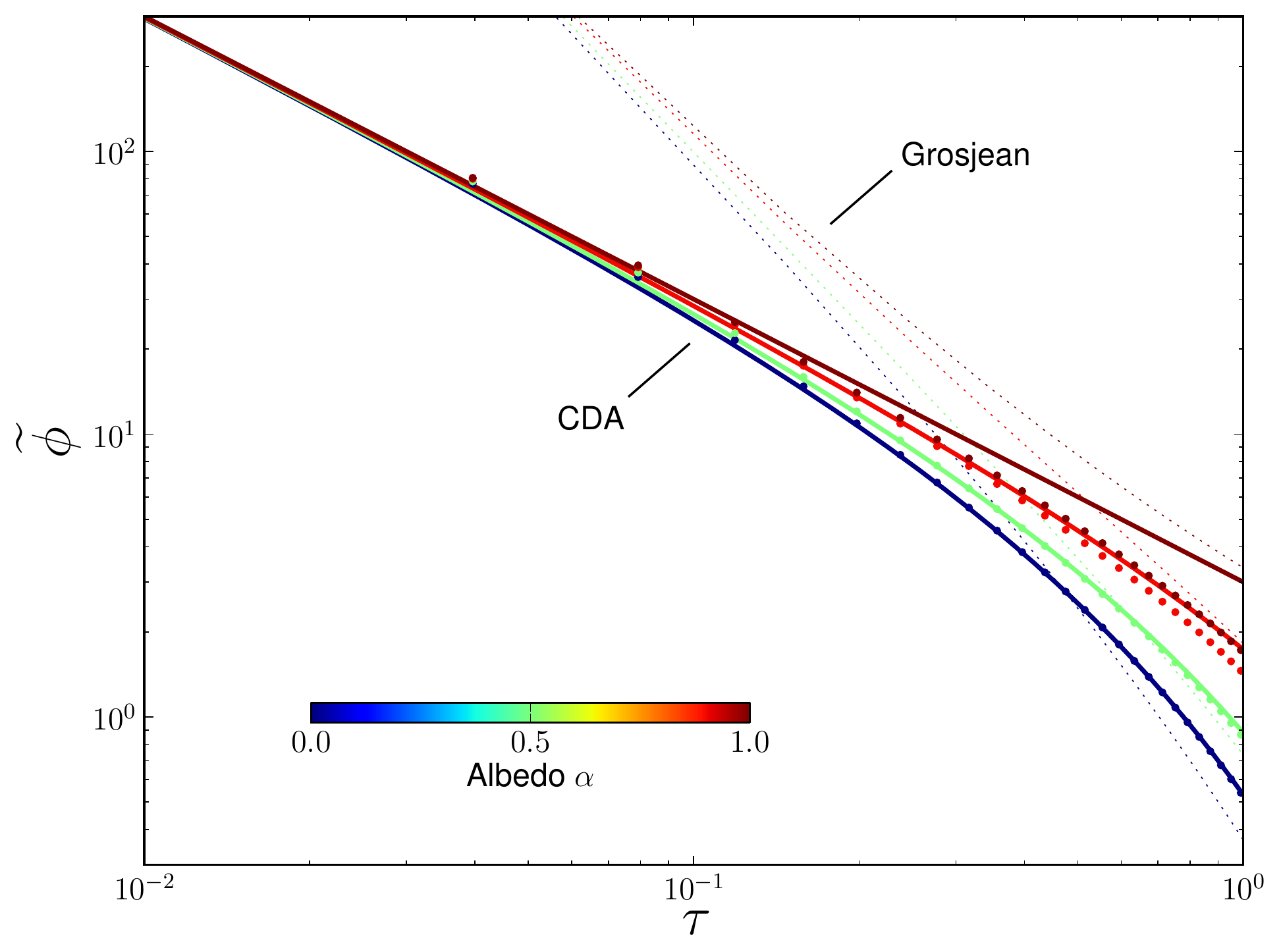}}
\subfigure[FLD\label{subfig:numerical_fld}]{\includegraphics[width=0.32\linewidth]{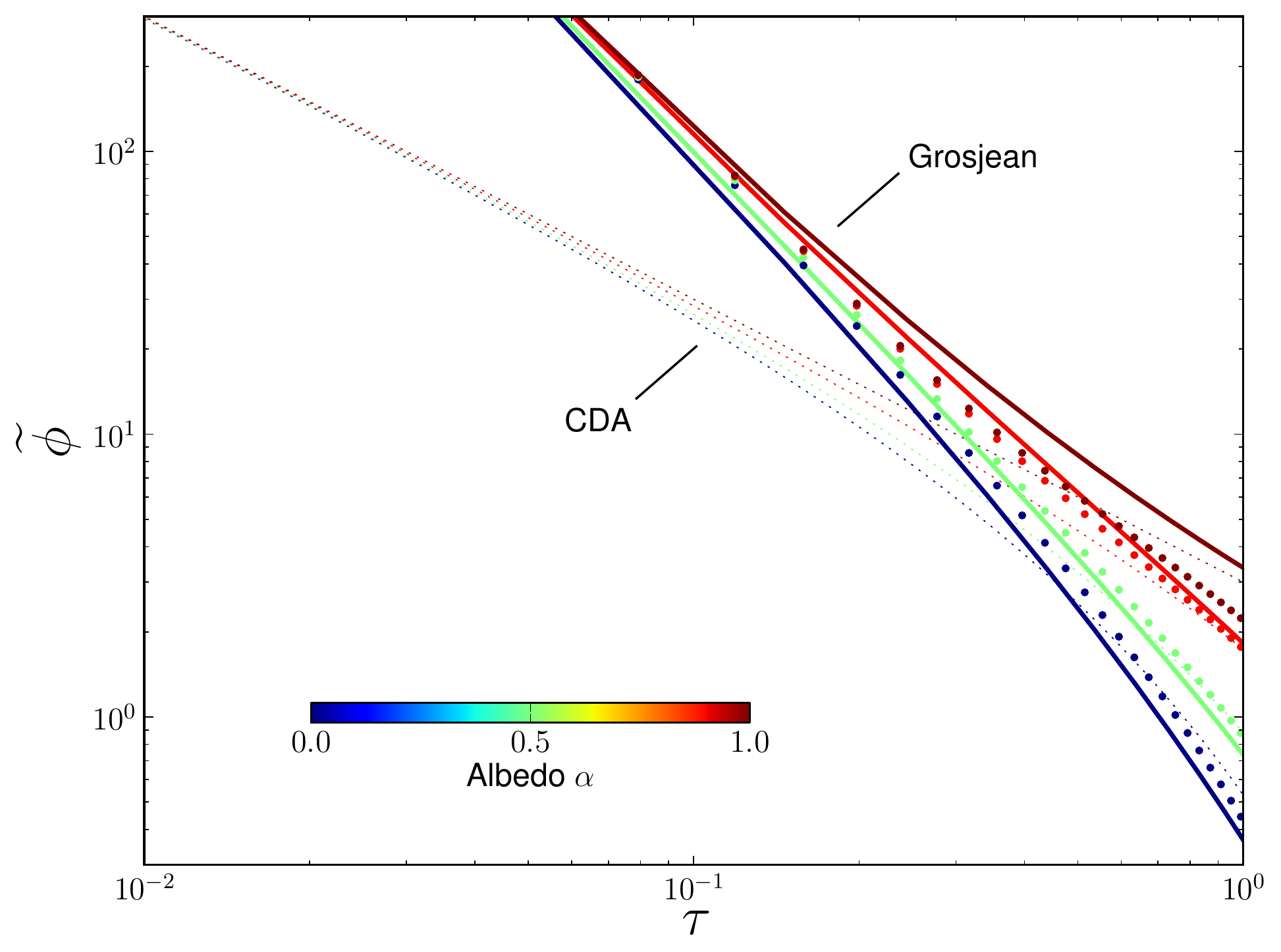}}
\hspace*{\fill}
\caption{Numerical solution for point source normalized fluence $\widetilde{\phi}$ (defined in Eqn.~\ref{eq:normalized_grosjean_fluence}) for various albedos. In \subref{subfig:numerical_cda2} and \subref{subfig:numerical_cda} we used the CDA solver, and in \subref{subfig:numerical_fld} we used our FLD solver. In all figures, points indicate the numerical solution for $\widetilde{\phi}$ at given optical depth $\tau$, and solid/dotted curves show the CDA theory Eqn.~(\ref{eq:cda_greensfunction}) and Grosjean transport theory Eqn.~(\ref{eq:grosjean}) as labelled. In \subref{subfig:numerical_cda2} we show the numerical solution of CDA with Dirichlet boundary conditions set from the analytical solution. It closely matches the CDA theory, demonstrating that the deviation of the numerical solve from the theory in~\subref{subfig:numerical_cda} is entirely due to the effect of the $\phi=0$ Dirichlet boundary condition applied in this case. In \subref{subfig:numerical_fld} we use the $\phi=0$ boundary condition.
}
  \label{fig:pointsource_validation}
\end{figure*}

To improve the rate of convergence, we use the successive over-relaxation (SOR) technique. Defining the
over-relaxation parameter $\omega$, where $0<\omega<2$, the update of $\phi_p$ at each stencil is modified to 
\begin{equation} \label{sor_phi_update}
\phi_p \leftarrow \omega \,\phi'_p + (1-\omega)\phi_p 
\end{equation}
where $\phi'_p$ is the updated value from Eqn.~(\ref{phi_update}). 
To monitor convergence, we compute a normalized residual as the RMS of the residual field $\bar{R}$ (ignoring voxels close to the boundary, to avoid boundary effects), divided by $\bar{j}$. Writing Eqn.~(\ref{phi_update}) as $\phi_p \leftarrow \mbox{numerator}/\mbox{denominator}$, we compute $R_p$ as:
\begin{equation}  \label{R_calc}
R_p = \left(\mbox{numerator} - \phi_p \cdot \mbox{denominator}\right) / \Delta l^2 \ .
\end{equation}
When $\bar{R}/\bar{j}$ falls below a suitable tolerance, e.g. $10^{-6}$ for double precision calculations, we regard the numerical solution as having converged. An overview of the complete process is given in Algorithm~\ref{redblack_sor_relaxation_algo}.

In Eqn.~(\ref{D_update}), we impose a lower bound on the extinction $\sigma_\epsilon$, as we found empirically that this tolerance has an effect on the convergence rate in the presence of vacuum regions. This effect is shown in Fig.~\ref{subfig:kt_min_wedge}, using our GPU solver applied to the Nebulae dataset. A value of $\sigma_\epsilon=10^{-3}/L$ (where $L$ is the grid extent), corresponding to a maximum optical depth of $1000$ box lengths, was sufficient to ensure good convergence with all of our test datasets. 

We implemented the numerical method described in both a testbed solver in double precision on the CPU, and a high performance GPU solver implemented in single precision via the NVIDIA CUDA framework.
In Figures~\ref{subfig:residual-against_sor} and~\ref{subfig:residual-against_iterations} we show the convergence properties of our CPU FLD solver on a simple test case where the extinction field defines an embedded spherical cloud (with a vacuum exterior) and an internal emission field generated via Perlin noise. These figures demonstrate that our numerical method converges over a broad range of SOR parameters.

%
%

\section{Results}
\label{sec:results}

In this section we apply our numerical FLD solver first to the point source problem discussed in Section~\ref{sec:point_source_theory}, and then to various heterogeneous datasets.

\subsection{Point Source in Homogeneous Medium}

We first applied our FLD solver to the point source problem presented in Section~\ref{sec:point_source_theory} to verify that it behaves according to the theory. The point source was modeled as a single emitting voxel at the center of the numerical grid, which had a resolution of $127^3$, and a homogeneous extinction corresponding to $\tau=4$ across the width of the grid. We set $j$ within the emitting voxel to the reciprocal of the voxel volume in order to correctly discretize a unit power point light. 
We ran both the CDA and FLD solvers (where the CDA solver is simply the FLD solver with flux limiter set to constant $\mathcal{F}=1/3$) for a range of albedos, and then extracted the normalized fluence as a function of radial distance.

In Fig.~\ref{fig:pointsource_validation}, we compare our numerical results with the analytical solutions.
Fig.~\ref{subfig:numerical_cda} shows the solution obtained using CDA, and Fig.~\ref{subfig:numerical_fld} shows the solutions obtained via FLD using the Levermore--Pomraning flux limiter. 
With the CDA solver, at low to medium albedo we obtain a good match between classical diffusion theory and the numerics, as expected since the CDA numerical solver is based on CDA theory. At high albedo, the effect of the boundaries becomes more pronounced, making the obtained CDA solution deviate from the CDA theory of the case of an infinite homogeneous medium.
However, we do not expect a perfect match between the FLD numerical solver and the Grosjean theory, since FLD is merely an approximation to the full transport theory.  Despite this, with the FLD solver we obtain a rough match to the Grosjean ``ballistic'' behavior at low optical depth, and the Levermore--Pomraning flux limiter seems to produce a reasonable quality match across the whole albedo range (the other flux limiters produce a similar though slightly worse match).
This indicates that the FLD approximation of the transition to the transport regime is working reasonably well.

\subsection{Practical Examples}

In this section, we compare renderings of heterogeneous datasets obtained via FLD, CDA, and a volumetric path tracer which serves as  ground truth. 
The path tracer uses the standard Woodcock tracking method \cite{raab2008unbiased} 
and was validated using Chandrasekhar's analytical solution for multiple scattering in a slab geometry \cite{chandrasekhar1960radiative}. 

\begin{figure}[!lb]
\setcounter{subfigure}{0}
\centering
\subfigure[Original resolution]{\includegraphics[width=0.49\linewidth]{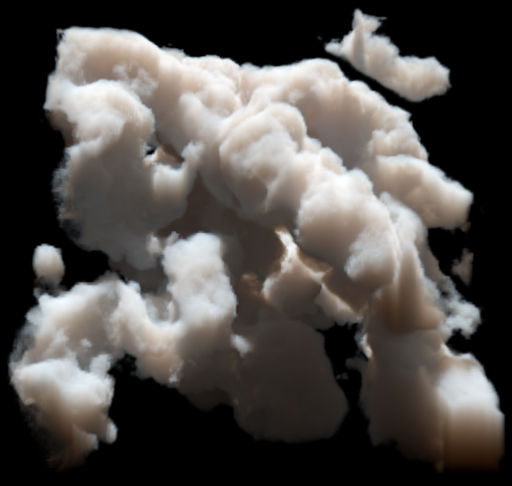}}
\subfigure[Quarter resolution]{\includegraphics[width=0.49\linewidth]{nebulae_blackbackground_fld_lowres.png}}
\caption{Comparison of our method when solving for multiply-scattered light using FLD at original resolution $200^3$ (left) and at resolution $50^3$ (right).}
\label{fig:res_compare}
\end{figure}

\begin{figure*}[!bth]
\includegraphics[width=1.0\linewidth]{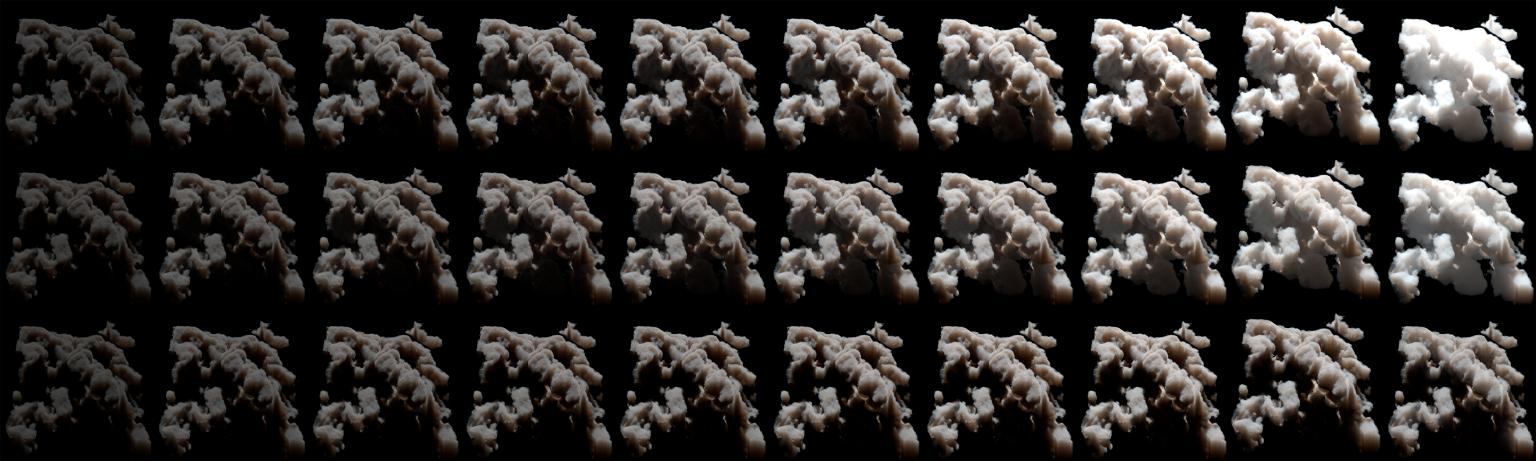}
\caption{Nebulae dataset rendered for albedo values from 0.1 (left) to 1.0 (right). The path-traced reference (top row) is compared against FLD (middle row) and CDA (bottom row).}
\label{fig:results_cloud_wedge}
\end{figure*}

In all cases we illuminated the volumes with a single directional external light source $L_l$.
For the FLD and CDA tests, the external lighting produces the multiple scattering source term $q_{ri}(\mathbf{x}) = L_l \sigma_s(\mathbf{x}) T(\mathbf{x})$, which we computed via raymarching and stored at the full resolution of the dataset extinction field.
We solve Eqn.~(\ref{fld_radiative_diffusion_equation}) for the fluence $\phi(\mathbf{x})$ using Algorithm~\ref{redblack_sor_relaxation_algo}, with source term $q_{ri}(\mathbf{x})$. We use the Levermore-Pomraning flux-limiter in all examples (the other flux-limiters were tried and produced very similar results). Then via Eqn.~(\ref{source_radiance_in_terms_of_phi}) we complete the primary raymarch Eqn.~(\ref{primary_raymarch}) to render the image.
Since multiply-scattered light tends to reduce higher-order angular moments, $\phi(\mathbf{x})$ has a much lower spatial frequency than the extinction field and therefore resolving it at the same resolution is not necessary. In fact, a major benefit of the diffusion approach is the ability to decouple the computation of the multiply-scattered light from the singly-scattered light, and solve for the multiply-scattered contribution more efficiently. To exploit this we therefore, in all our examples, solved for $\phi(\mathbf{x})$ on grids with a quarter of their original resolution. Fig.~\ref{fig:res_compare} demonstrates that the effect of solving for the multiply-scattered light on such a coarser grid has a low visual impact. 

\begin{figure*}[!tb]
\hspace*{\fill}%
\subfigure[Path tracing (\unit{19}{\minute})]{\includegraphics[width=0.3\linewidth]{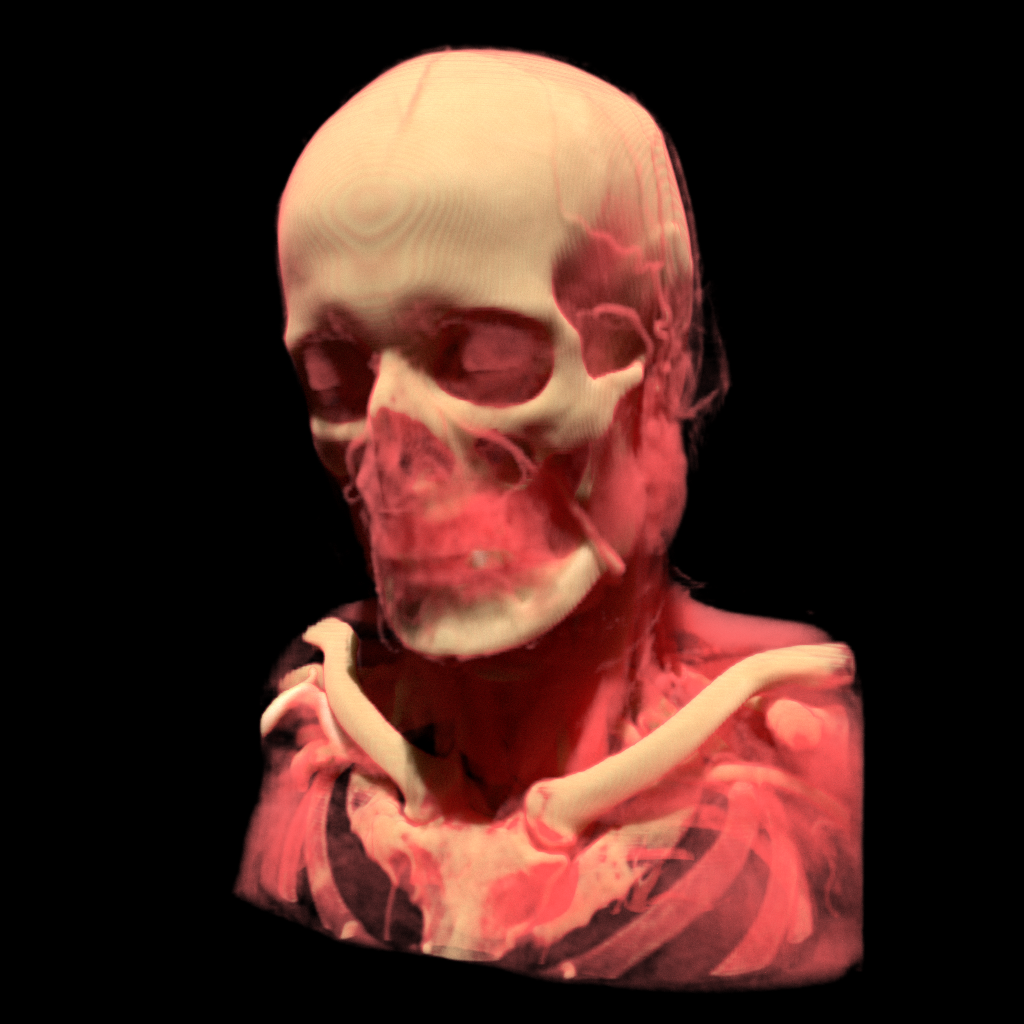}}
\hspace{1.5pt}
\subfigure[FLD (\unit{1.8}{\second})]{\includegraphics[width=0.3\linewidth]{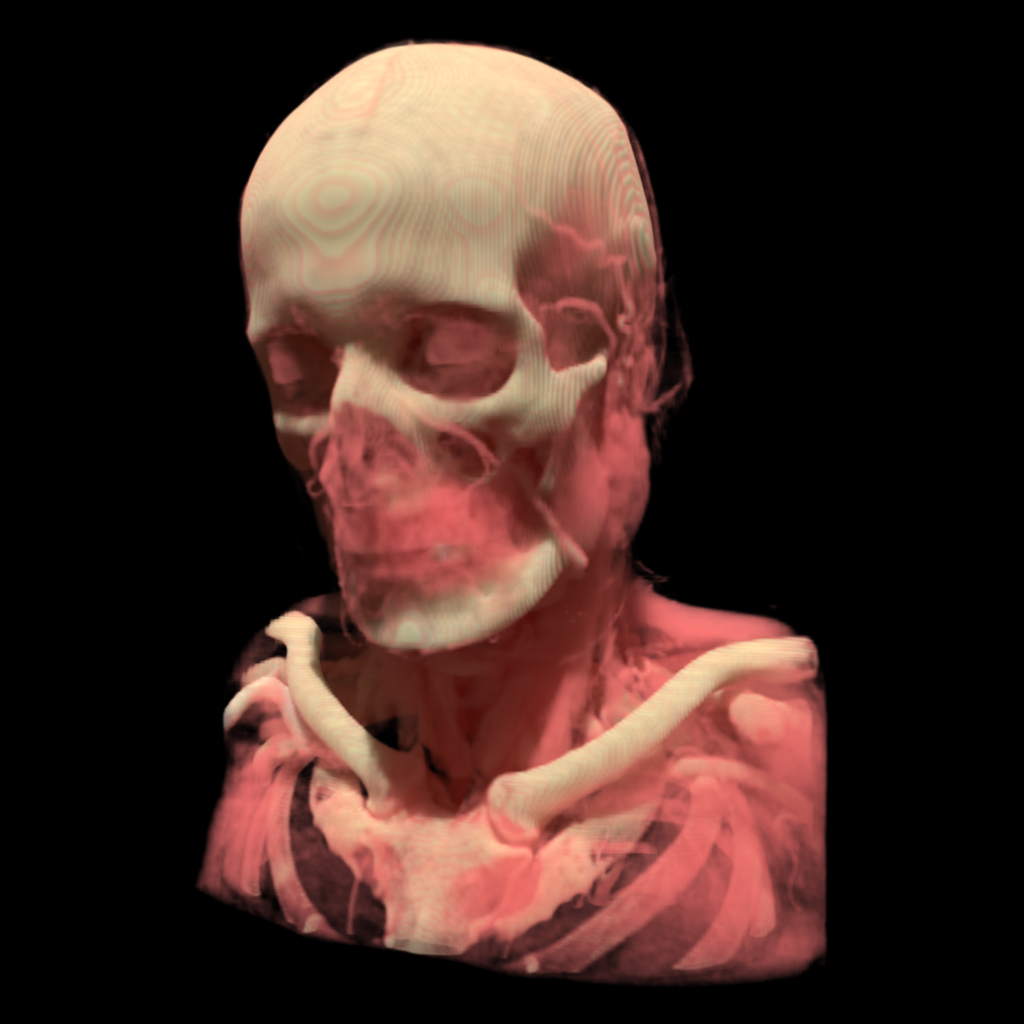}}
\hspace{1.5pt}
\subfigure[CDA (\unit{1.3}{\second})]{\includegraphics[width=0.3\linewidth]{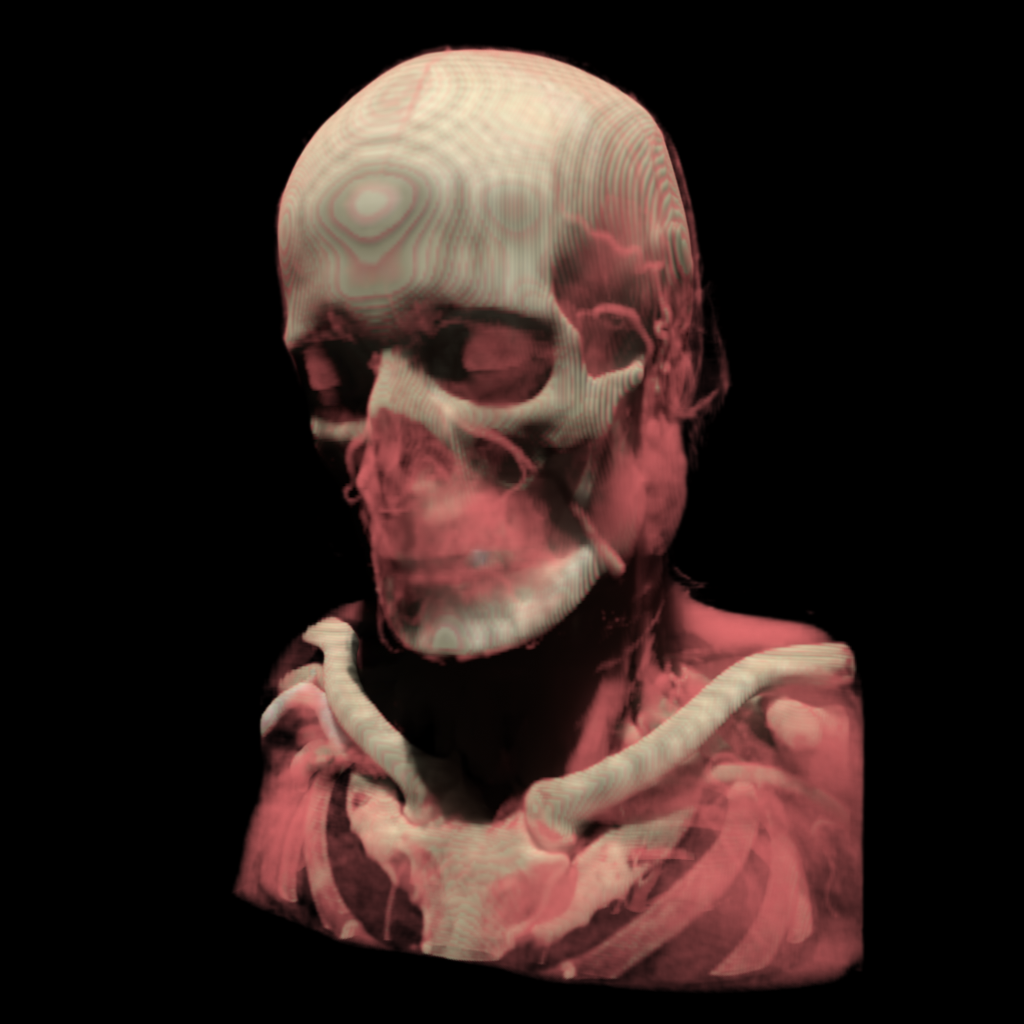}}%
\hspace*{\fill}%
\caption{MANIX CT-scan dataset with spatially varying albedo. Extinction coefficient and albedo are driven by transfer functions. Note how FLD better captures indirect light at the neck in comparison to CDA.}
\label{fig:results_manix_tissue}
\end{figure*}

We implemented all three approaches in single precision using the NVIDIA CUDA framework.
When measuring performance, we included in the CDA and FLD cases the time taken for primary raymarching and generation of the source term (light baking), as well as the time taken during the diffusion solve. All results were obtained with a NVIDIA GTX TITAN GPU.
We ran the diffusion solver until convergence at the level of $\bar{R}/\bar{j}=10^{-4}$, which is the practical limit for the single precision computation. With all our path-traced images, we used $1000$ Monte Carlo samples per-pixel, which we found to strike a good balance between image quality and computation time.

All images were rendered at a resolution of $1024^2$ pixels, and tone-mapped using standard gamma compression. For each dataset, exposure was adjusted for ground truth rendering to avoid blown highlights and then the same tone mapping was used for FLD and CDA.

\begin{figure*}[!tb]
\hspace*{\fill}%
\subfigure[Path tracing (\unit{21}{\minute})]{\includegraphics[width=0.3\linewidth]{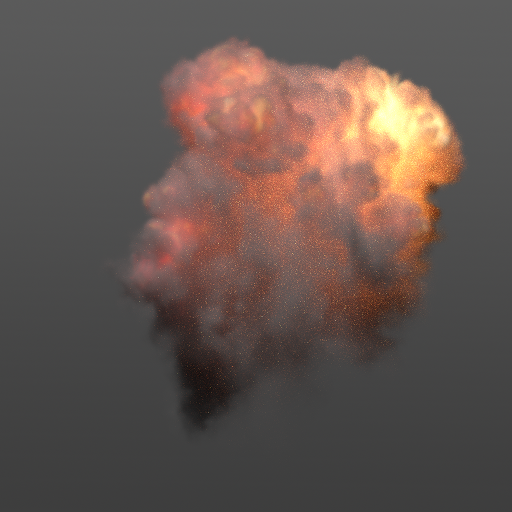}}
\hspace{1.5pt}
\subfigure[FLD (\unit{0.7}{\second})]{\includegraphics[width=0.3\linewidth]{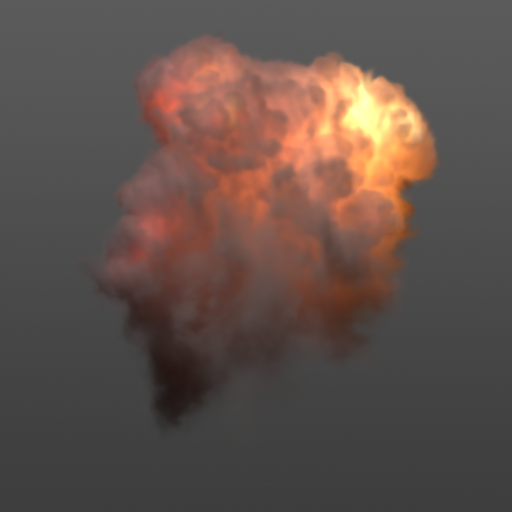}}
\hspace{1.5pt}
\subfigure[CDA (\unit{0.5}{\second})]{\includegraphics[width=0.3\linewidth]{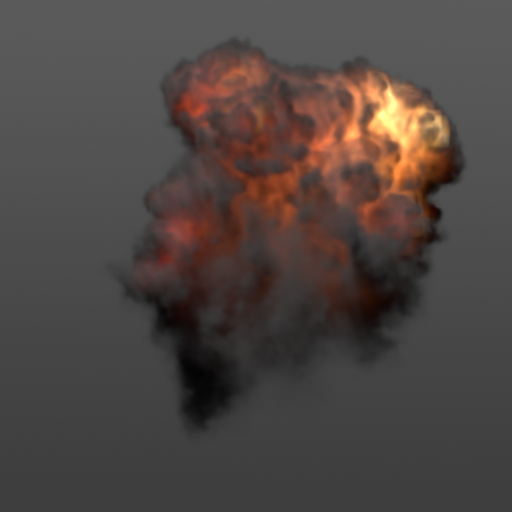}}%
\hspace*{\fill}%
\caption{Fluid simulation dataset lit by internal blackbody emission and an external light source.}
\label{fig:results_explosion}
\end{figure*}

The Nebulae example, shown in Fig.~\ref{fig:nebulae}, was generated from a noise-displaced implicit surface. Color is introduced by a color-dependent scaling of the extinction field. The resolution of the dataset is $200^3$ and the albedo is $0.9$. Note the differences of the shadowed regions and the volume blob at the bottom of the dataset, which only receives indirect light coming from other illuminated parts of the volume. While CDA is completely black in those regions, FLD finds a solution which is much closer to the path-traced result. To compare the behavior of both methods as albedo is varied, we show in Fig.~\ref{fig:results_cloud_wedge} that the rendering of the Nebulae dataset with FLD produces better results than CDA over the whole range of albedos, though with a more pronounced difference at high albedos.

The example in Fig.~\ref{fig:results_manix_tissue} is based on a CT-scan dataset, which has a resolution of $256{\times}329{\times}256$. In this example, the density field in the original dataset is mapped to a per-channel albedo field and a (color channel independent) extinction field via a transfer function which colors tissue and bone areas distinctly. The image shows that FLD better captures indirect light at the neck and the translucent nature of tissue in the path-traced rendering, when compared to CDA. 

The example in Fig.~\ref{fig:results_explosion}  
is based on a fluid simulation with a resolution of $201{\times}272{\times}230$. There is an external light source and a constant albedo of $1.0$. In addition, there is a black body emission field which is calculated from the temperature field of the simulation. Again FLD matches the ground truth closer than CDA.

These examples demonstrate that FLD produces results which are closer to path tracing than CDA for a variety of datasets and parameterizations. Significant improvements over CDA appear especially in situations with indirect lighting or transitions from highly dense media to regions of low density or vacuum. As a modified diffusion theory, FLD of course still deviates to a certain extent from the true RTE solution, as all diffusion methods throw away all but the lowest angular moments of the radiation field.
Our performance measurements indicate that FLD is roughly twice as computationally expensive as CDA, but is still very fast when compared with path tracing. We noted that in our GPU implementation, this loss in performance is dominated by the additional memory access per voxel which is required to compute the Knudsen number.

\section{Conclusion}
\label{sec:conclusion}

We introduced the flux-limited diffusion approximation for multiple scattering in participating media. Flux-limited diffusion originated in the field of astrophysics as an improvement to CDA which produces results of higher accuracy when applied to heterogeneous media containing regions of low extinction or vacuum. While FLD is slightly more computationally expensive than CDA, it is still much faster than path tracing. We provided a numerical method which we checked for convergence and validated using an analytical solution for scattering of a point source. We rendered various heterogeneous volumetric datasets with FLD, CDA, and path tracing, and demonstrated that FLD produces a better match to path-traced ground truth than CDA.

For future work, we suggest that the extension of FLD to anisotropic phase functions may be of particular importance. In the anisotropic radiative diffusion theory, the extinction is replaced by a ``reduced extinction'', $\sigma_t^{\prime} = \sigma_t(1-\alpha g)$, where $g\in[0,1]$ is the phase function anisotropy factor \cite{stam1995multiple, jensen2001practical}. It causes a divergence in the diffusion coefficient, and correspondingly the flux, as $g\rightarrow 1$ (strong forward scattering) and $\alpha\rightarrow 1$ (high albedo), even in optically thick media. Thus, in the case of strongly forward scattering high-albedo media, flux-limiting is an essential ingredient to obtain physically accurate results from diffusion methods. This is of practical importance in the application of diffusion-based methods to, for example, the rendering of Mie scattering \cite{Bouthors:2008:IMA:1342250.1342277, Yue:2010:UAS:1882261.1866199} in atmospheric clouds. 
The form of the flux is also different from Eqn.~(\ref{phi_and_E}) in the presence of anisotropy, with an additional term involving the first moment of the external flux \cite{stam1995multiple, DBLP:conf/wscg/MaxSMIN04}, which suggests that a detailed theoretical analysis is needed to understand how to achieve flux limiting in this case.

There might also be substantial performance benefit to be gained by developing a more sophisticated numerical algorithm for solving the flux-limited diffusion equation than the one presented here, such as a non-linear multigrid solver \cite{Briggs:2000:MT:357695}.
Finally, it is worth noting that any existing graphics technique which is based on classical diffusion theory, such as the work of Arbree et al. \cite{Arbree:2011:HSS:1990770.1990989},  could be easily augmented with the FLD technique for improved accuracy.

\vspace{-0.3cm}

\section{Acknowledgements}
This work was partially funded by the graduate scholarship program Digital Media, granted by the state Baden-W{\"u}rttemberg, Germany, and the Cluster of Excellence in Simulation Technology (EXC 310/1), granted by Deutsche Forschungsgemeinschaft (DFG). The authors would like to thank Jose Esteve and Sebastian Herholz for helpful suggestions.

\bibliographystyle{eg-alpha}
\bibliography{main}

\end{document}